\documentclass[%
reprint,
amsmath,amssymb,
aps,
prb,
]{revtex4-1}

\usepackage{xcolor}
\usepackage{graphicx}% Include figure files
\usepackage{dcolumn}% Align table columns on decimal point
\usepackage{bm}% bold math
\begin{document}
	
\title{Slow quenches in Chern insulator ribbons}

\author{ Lara Ul\v{c}akar}

\affiliation{Jozef Stefan Institute, Jamova 39, Ljubljana, Slovenia}

\email{lara.ulcakar@ijs.si}

\author{Jernej Mravlje}

\affiliation{Jozef Stefan Institute, Jamova 39, Ljubljana, Slovenia}

\author{Toma\v{z} Rejec}

\affiliation{Jozef Stefan Institute, Jamova 39, Ljubljana, Slovenia}

\affiliation{Faculty for Mathematics and Physics, University of Ljubljana, Jadranska
	19, Ljubljana, Slovenia}

\date{\today}	

\begin{abstract}
	
	We investigate slow quenches in Chern insulators in ribbon geometry. We consider the Qi-Wu-Zhang model and slowly ramp the parameters (large time of the quench $\tau$) from a non-topological (Chern number = 0) to a topological regime (Chern number $\ne$ 0). In contrast to the Haldane model considered in [L. Privitera and G. E. Santoro, Phys. Rev. B 93, 241406(R) (2016)] earlier, the in-gap state degeneracy point is pinned to an inversion symmetric momentum, which changes the behavior drastically.  The density of excitations in the in-gap states scales with the quench time as $\tau^{-1/2}$ as the ramp becomes slow, and the Kibble-Zurek mechanism applies. Despite the slower scaling of the density of in-gap excitations with $\tau$, the Hall conductance after the quench deviates from that of the ground state of the final Hamiltonian by an amount that drops as $\tau^{-1}$.

\end{abstract}

\maketitle

\section{Introduction}

Topological insulators have attracted a lot of attention in recent years due to their novel properties such as quantum (spin) Hall effect in the absence of magnetic field.\cite{Hasan10,Qi11} Such systems can host in-gap conducting edge states that are robust against disorder and account for the exact quantization of the Hall conductance. The existence of the edge states is connected to the non-trivial topological order parameter of the bulk. Several experimental realizations of topological insulators in cold atom systems \cite{Tarruell12,Aidelsburger13,Miyake13,Goldman13,Dauphin13,Jotzu14,Wu16,Flaschner16} and in solid state systems \cite{HgTeExp,Yu10,InAsExp,Chang13,Bismuthene,Tokura19} have been recently demonstrated.

Response of topological insulators to quantum quenches between different topological phases has been under an active investigation. Studies of bulk systems considered sudden (a parameter of the Hamiltonian is rapidly changed) and slow quenches (a parameter of the Hamiltonian is varied smoothly). Sudden quenches result in a non-quantized value of the Hall conductivity. \cite{Caio2016,Unal16} Refs.~\onlinecite{zoller,Ulcakar2018} considered slow quenches and showed that the (spin) Hall conductivity approaches the quantized value of the final ground state. The ground-state value is never fully reached since during the quench the energy gap closes and the system always gets excited. The number of excitations \cite{Damski05,Dutta10,Ulcakar2018} follows the Kibble-Zurek (KZ) argument, \cite{Kibble,Zurek} which predicts that the density of excitations scales with the quench time to the power given by critical exponents, associated with the critical point across which the system is quenched. Non-equilibrium classification of topological insulators is presented in Ref.~\onlinecite{McGinley18}. In the case of two-dimensional Chern insulators, the Chern number remains well defined and is equal to the one evaluated in the initial state. \cite{Rigol2015,Caio2015,PxPySuperFluid,Ulcakar2018} Cold atoms could be used to study quantum quenches, where a time-dependent change of a parameter of a Hamiltonian could be performed by changing the external gauge fields in the optical lattice \cite{Tarruell12}.

Quenches in finite-sized systems with edges were explored less. Sudden quenches in such systems were studied in Refs.~\onlinecite{Caio2015,Duta17,Mitra16}, which showed that after a sudden quench from a topological to a trivial regime, edge currents relax into the interior of the system. Ref.~\onlinecite{Privitera16} studied slow quenches in Floquet Chern insulators in which Floquet Hamiltonian is approximately given by the Haldane model.\cite{HaldaneModel} The degeneracy point of the in-gap states moves in $k$-space during the quench, which results in a perfect population of selected in-gap states after the quench. As a consequence of this anomalous in-gap excitation production, the excitations in the system do not follow the KZ scaling. Deviations from the KZ scaling due to edge states were also observed in the one-dimensional Creutz ladder.\cite{Bermudez09,Bermudez10}

In this paper we study slow quenches in a Chern insulator with edges, represented by the Qi-Wu-Zhang (QWZ) model \cite{QWZmodel} in a ribbon geometry. After a quench to the topological side, the in-gap states appear and are occupied, in contrast to the bulk Chern number that remains equal to that in the initial Hamiltonian.\cite{Rigol2015,Caio2015,Ulcakar2018} The number of generated excitations decreases as the quench becomes slow. The number of the excitations in the in-gap states scales with the quench time $\tau$ as $\tau^{-1/2}$ whereas the number of excitations in the bulk states scales as $\tau^{-1}$. The scaling of the in-gap and the bulk excitations is well described by the KZ mechanism. These results differ from the ones in Ref.~\onlinecite{Privitera16} because in our case the degeneracy point is pinned to an inversion symmetric momentum due to the inversion symmetry.  We also calculated the Hall conductance, which approaches the ground-state value of the final Hamiltonian. The deviation of the post-quench Hall conductance from the quantized value arises both due to bulk and in-gap excitations. Both bulk and in-gap contributions scale as $\tau^{-1}$, which in turn is the same as also for the ``bulk-only'' Chern insulator with periodic boundary conditions in both directions.

The paper is structured as follows: In Sec.~\ref{sec:model} we introduce the QWZ model.  In Sec.~\ref{sec:numExc} we turn to the state of the system after a quench and discuss the scaling of  the number of excitations in the bulk and in the in-gap bands with the duration of the quench. In Sec.~\ref{sec:HallCond} we consider the Hall response after the quench and in Sec.~\ref{sec:breaking} and Sec.~\ref{sec:breakchiral} we discuss effects of the breaking of the inversion symmetry and the chiral symmetry on the system after a quench. In Appendix~\ref{app:GSelectric} we pedagogically discuss the linear response of the system in the ground state to the electric potential. In Appendix~\ref{app:bandProp} we derive the scaling of the total number of the in-gap excitations, followed by the calculation of critical exponents of the in-gap bands in Appendix~\ref{app:critExp}. In Appendix~\ref{app:HallDer} we evaluate the time-averaged Hall conductance for systems out of equilibrium using the time-dependent perturbation theory.

\section{The model}
\label{sec:model}
We study a Chern insulator described by the QWZ model,\cite{QWZmodel} also known as the "half BHZ model" as it is the basic building block of the Bernevig-Hughes-Zhang model\cite{BHZmodel} for the quantum spin Hall effect. The real-space QWZ Hamiltonian reads
\begin{equation}
\begin{split}
\hat H=\sum_{x,y}( |x,y\rangle\langle x,y|\otimes u\hat\sigma_z+
|x+1,y\rangle\langle x,y|\otimes\hat t_x+\\
\textrm{h.c.}+|x,y+1\rangle\langle x,y|\otimes\hat t_y+\textrm{h.c.}),\label{eq:QWZreal}
\end{split}
\end{equation}
where $x$ and $y$ index unit cells of a square lattice and Pauli matrices $\hat\sigma_i$, $i\in\{x,y,z\}$ represent the internal degree of freedom due to a presence of two orbitals $|A\rangle$ and $|B\rangle$ per unit cell. The Hamiltonian contains staggered orbital binding energy $u$ and nearest neighbour hoppings $\hat t_x=\frac{\hat\sigma_z+i\hat\sigma_x}{2}$ and $\hat t_y=\frac{\hat \sigma_z+i\hat \sigma_y}{2}$ in $x$ and $y$ directions, respectively.

The system with periodic boundary conditions has the Hamiltonian $\hat H=\sum_{k_x,k_y}|k_x,k_y\rangle\langle k_x,k_y|\otimes\hat H(k_x,k_y)$ with
\begin{equation}
\begin{split}
\hat{H}(k_x,k_y)=(u+\cos{k_x}+\cos{k_y})\hat{\sigma}_z+
\sin{k_x}\hat{\sigma}_x+\sin{k_y}\hat{\sigma}_y,\label{eq:Hbulk}
\end{split}
\end{equation}
where $k_x$ and $k_y$ are the Cartesian components of the momentum vector $\mathbf{k}$. Throughout the manuscript the lattice constant is fixed to 1. The energy gap between the valence and the conduction band as well as the topological phase of the system are controlled by the parameter $u$. The gap closes at critical values $u_c\in\{-2,0,2\}$ where the topological invariant -- the Chern number $C$ -- of the ground state changes its value:
\begin{equation}
C=
\begin{cases}
1, &\quad 0<u<2, \\
-1, &\quad -2<u<0,\\
0, &\quad |u|>2. \\
\end{cases}\label{eq:chernN}
\end{equation}
At a critical value of $u$ the two bands form a Dirac cone. The Hamiltonian Eq.~\eqref{eq:Hbulk} possesses the inversion symmetry, 
\begin{equation}
\hat H(-k_x,-k_y)=\hat \sigma_z\hat H(k_x,k_y)\hat\sigma_z.
\end{equation} 
Additionally, the Hamiltonian has also the following combination of the reflection and the chiral symmetry:
\begin{equation}
\hat H(-k_x,k_y)=-\hat \sigma_x\hat H(k_x,k_y)\hat\sigma_x.
\label{bulkchiral} 
\end{equation}

\begin{figure}[h]
	\centering
	{\includegraphics[width=125pt]{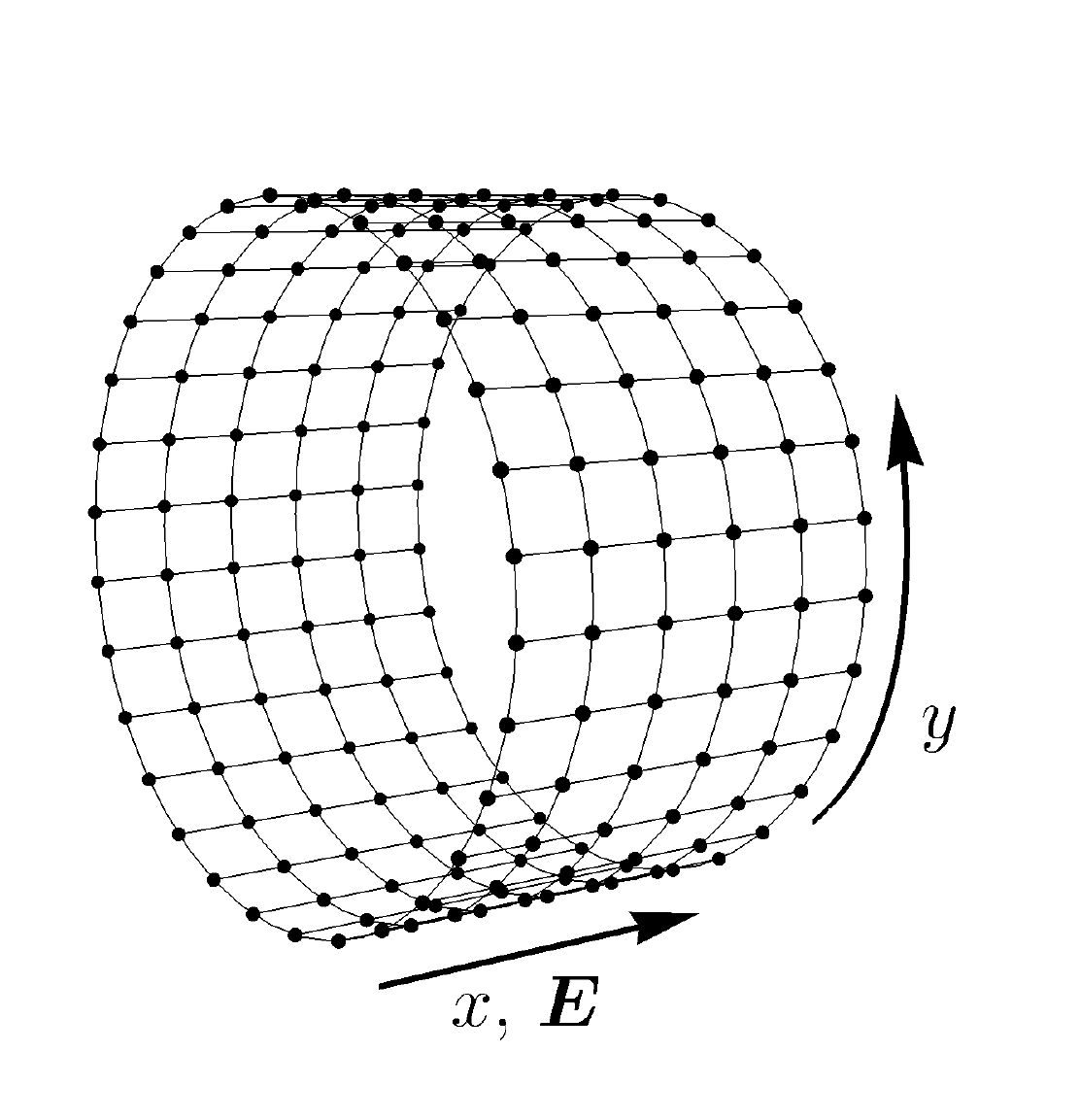}}
	
	\caption{Ribbon geometry with periodic and open boundary conditions along $y$- and $x$-directions, respectively. Homogeneous electric field $\boldsymbol E$ is applied in $x$-direction to probe the Hall conductance.}
	\label{fig:ribbon}
\end{figure}

In this work we study the QWZ model in the ribbon geometry where periodic boundary conditions are imposed along $y$-direction and open boundary conditions along $x$-direction, as presented in Fig.~\ref{fig:ribbon}. 
Due to the translation invariance in $y$-direction it is convenient to use the basis $|k_y\rangle\otimes|x\rangle$, where  $k_y\in[-\pi,\pi)$ is a Bloch wave vector and $x\in\{1,\dots,N_x\}$ is a lattice site in $x$-direction. In this basis the QWZ Hamiltonian is block-diagonal, $\hat{H}=\sum_{k_y}|k_y\rangle\langle k_y|\otimes\hat{H}(k_y)$ with
\begin{equation}
\begin{split}
\hat{H}(k_y)=\sum_{x=1}^{N_x-1}|x+1\rangle\langle x|\otimes\hat t_x+\mathrm{h.c.}+\\
\sum_{x=1}^{N_x}|x\rangle\langle x|\otimes\left(\left(\cos k_y+u\right)\hat{\sigma}_z+\sin k_y \hat{\sigma}_y\right).\label{eq:H}
\end{split}
\end{equation}
An electron occupying the $n$-th subband with the eigenenergy $\varepsilon_n(k_y)$ is described by the wave function $|\Psi_n(k_y)\rangle=|k_y\rangle\otimes|u_n(k_y)\rangle$, where $|u_n(k_y)\rangle$ is an eigenstate of $\hat{H}(k_y)$.

In a topologically non-trivial phase a ribbon hosts chiral in-gap states that are localized at the edges and exhibit an exponentially small gap of the order $e^{-N_x/\xi}$, where $\xi$ is the localization length of the in-gap states. 
Due to the inversion symmetry of the ribbon Hamiltonian, $\hat{H}(k_y)=\hat{U}^\dagger\hat{H}(-k_y)\hat{U}$ with $\hat{U}=\hat{P}\otimes\hat{\sigma}_z$ and $\hat{P}|x\rangle=|N_x+1-x\rangle$, the avoided crossing is pinned to an inversion-symmetric momentum $k_y=0$ or $k_y=\pi$. Due to Eq.~\eqref{bulkchiral}, the ribbon Hamiltonian possesses also the chiral symmetry, $\hat{H}(k_y)=-\hat{U}^\dagger\hat{H}(k_y)\hat{U}$ with $\hat U=\hat P\otimes\hat\sigma_x$. This makes the energy spectrum symmetric and thus pins the crossing to zero energy. The effects of breaking of these symmetries will be studied in Secs.~\ref{sec:breaking} and \ref{sec:breakchiral}.

Neglecting the avoided crossing due to hybridization of the edge states, we assign indices $L$ and $R$ to in-gap subbands with a positive and with a negative slope, respectively, as the former are localized to the left ($x=1$) edge while the latter are localized to the right ($x=N_x$) edge near $k_y=0$. The energy spectrum of a ribbon with $N_x=20$ at $u=-1.2$ is shown in Fig.~\ref{fig:BandExc}(a).

We evaluate the Hall conductance $G_{yx}$ by calculating the current in $y$-direction,
\begin{equation}
\hat{I}_y=\frac{1}{N_y}\sum_{k_y}|k_y\rangle\langle k_y| \otimes q\partial_{k_y}\hat{H}(k_y),
\label{eq:CurrOp}
\end{equation}
as a response to turning on a homogeneous electric field in $x$-direction  adiabatically as $E_x(t)=E_0[1-\exp (-t/\tau_E)]$. Here $q$ is the electron charge and, throughout the manuscript, we set $\hbar=1$. In presence of the electric field, the ribbon Hamiltonian $\hat{H}(k_y)$ thus acquires an additional term $\hat{V}(t)=-q E_x(t)(\hat{x}-\frac{N_x+1}{2})$, with $\hat{x}=\sum_{x=1}^{N_x}x|x\rangle\langle x|$.

In order to investigate the distribution of the current across a ribbon we calculate the current flowing through a particular site \cite{Caio2015} at position $\mathbf r=(x, y)$,
\begin{equation}
\hat{\boldsymbol{j}}(\mathbf{r})=q\sum_{\mathbf{r}'}-\frac{i}{2}\boldsymbol{\delta}_{\mathbf{r}'\mathbf{r}}\left(|\mathbf{r}'\rangle\langle \mathbf{r}|\otimes \hat{t}_{\mathbf{r}'\mathbf{r}}-\mathrm{h.c.}\right),
\end{equation}
where  $\boldsymbol{\delta}_{\mathbf{r}'\mathbf{r}}$ is the vector pointing from site $\mathbf{r}$ to site $\mathbf{r}'$ and $\hat{t}_{\mathbf{r}'\mathbf{r}}$ is the corresponding hopping operator from Eq.~\eqref{eq:QWZreal}.

\section{Slow quench}
\label{sec:quench}
We perform quenches by smoothly varying the parameter $u$ as $u(t)=u_0+(u_1-u_0) \sin^2(\frac{\pi}{2}\frac{t}{\tau})$ for $t\in[0,\tau]$ from deep in the trivial regime at $u_0=-2.8$ to deep in the topological regime at $u_1=-1.2$. All the results are calculated for a ribbon of finite length of $N_y=201$.

\subsection{Number of excitations}
\label{sec:numExc}
\begin{figure}[h]
	{\includegraphics[width=125pt]{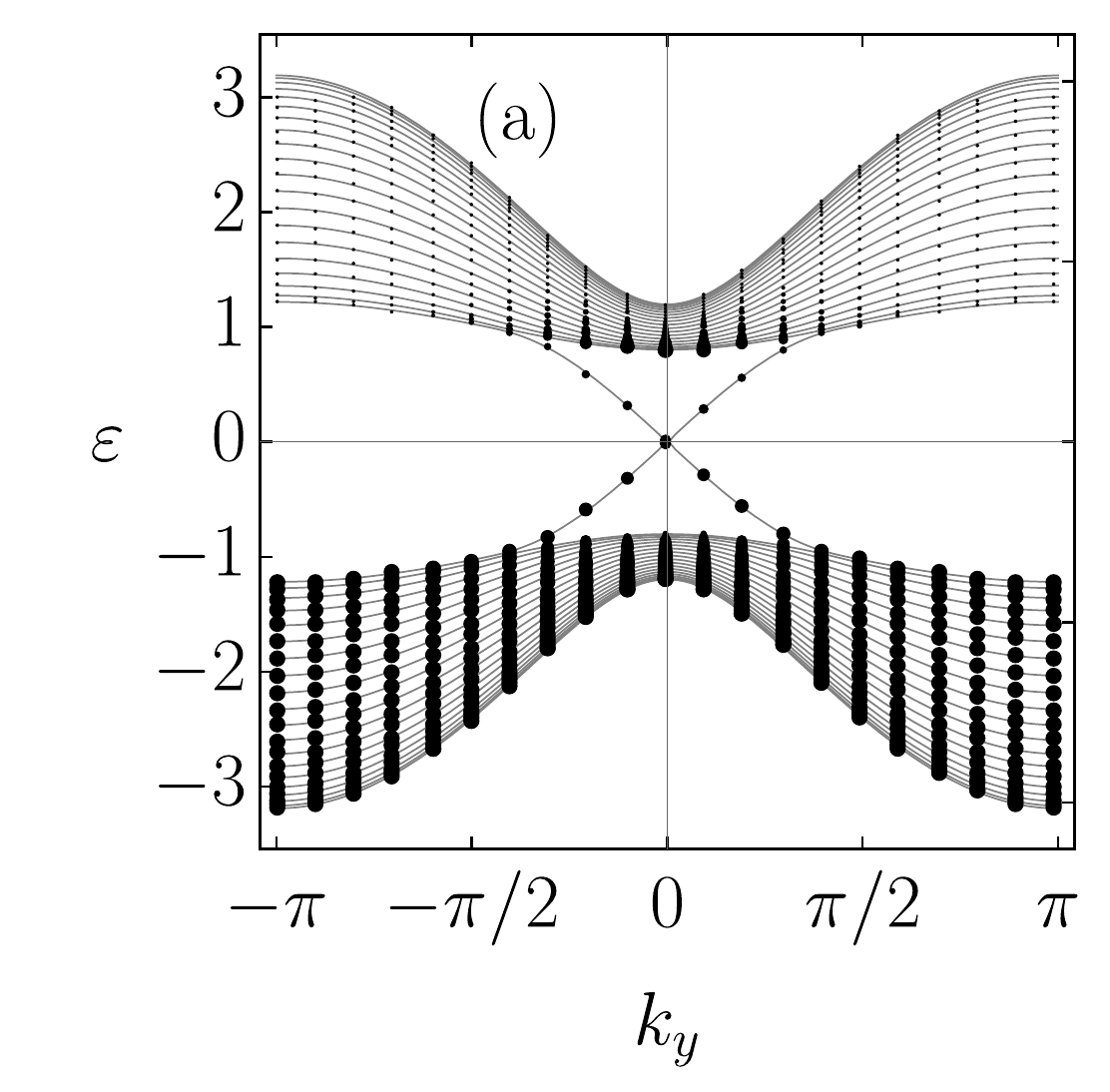}\includegraphics[width=125pt]{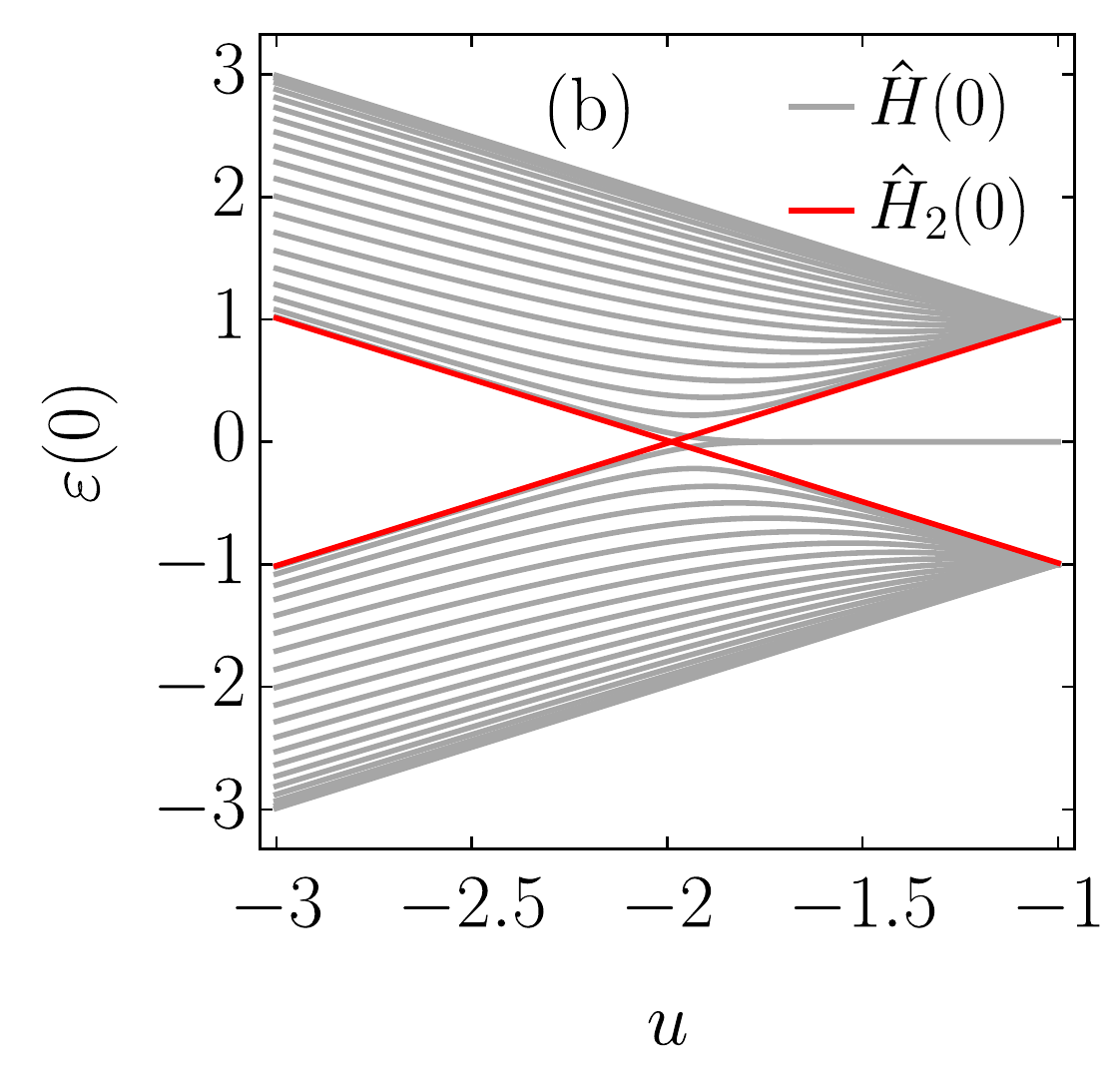}}
	
	\caption{ a) Occupancy of energy bands of the ribbon of width $N_x=20$ at $u=-1.2$ after the quench with $\tau=1$. The surface of a point is proportional to the probability of finding a particle in the corresponding eigenstate. (b) Energy levels of the whole Hamiltonian $\hat{H}(k_y)$ (gray) and of the two-level Hamiltonian $\hat{H}_2(k_y)$ (red) at $k_y=0$ as a function of $u$.}
	\label{fig:BandExc}
\end{figure}

\begin{figure}[h]
	\centering
	{\includegraphics[width=125pt]{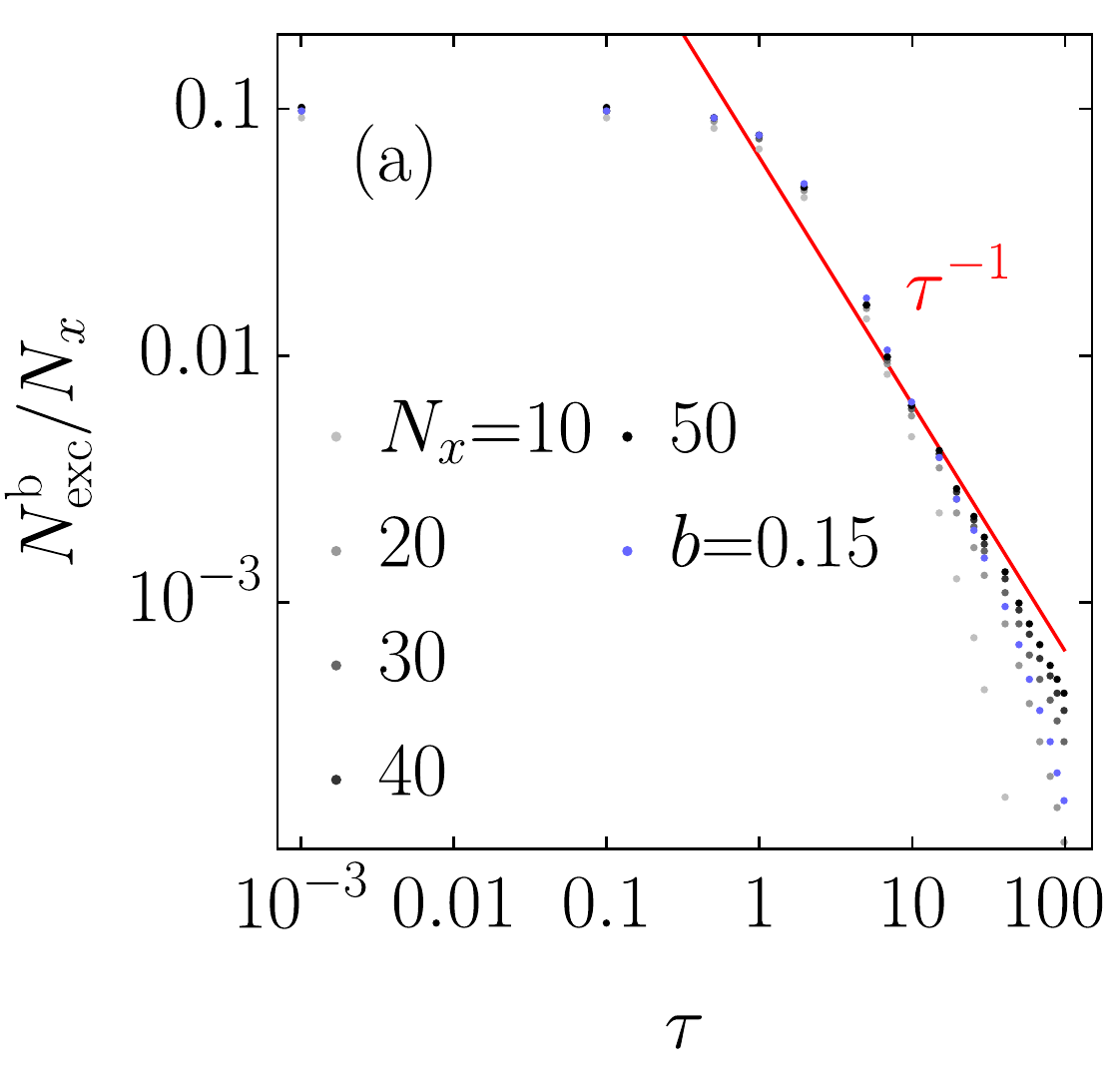}\includegraphics[width=125pt]{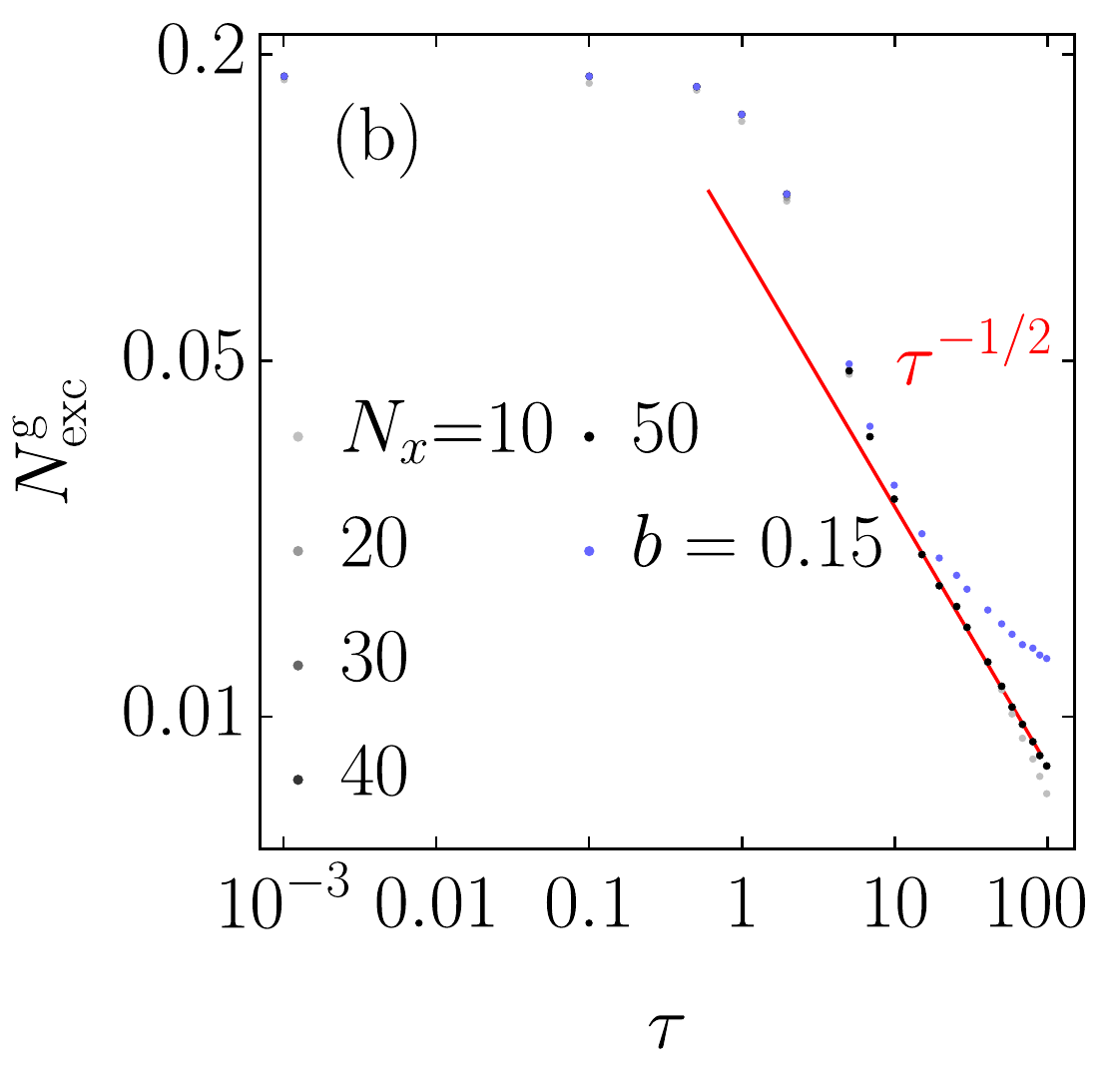}}
	\caption{ Number of excitations (a) in the bulk and (b) in the in-gap states as functions of $\tau$ for different ribbon widths, ranging from $N_x=10$ (light grey) to $N_x=50$ (black). Red lines denote (a) the exact result for the system with periodic boundary conditions in both directions which scales as $\tau^{-1}$ and (b) a fitted  $\tau^{-1/2}$ line. Blue dots show results for a Hamiltonian with a broken inversion symmetry of magnitude $b=0.15$ and $N_x=20$.}
	\label{fig:TotExc}
\end{figure}

\begin{figure}[h]
	\centering
	\includegraphics[width=125pt]{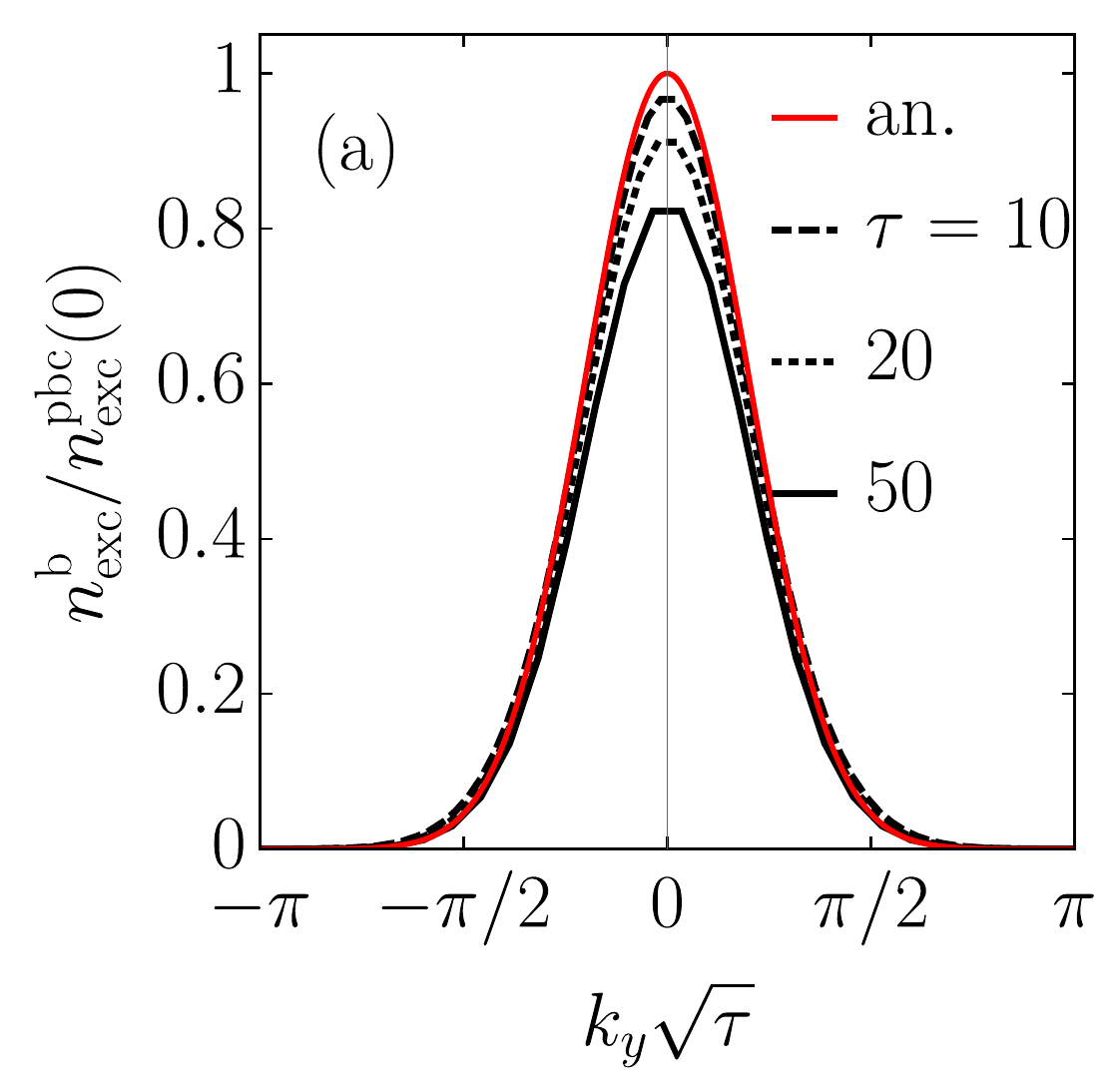}{\includegraphics[width=125pt]{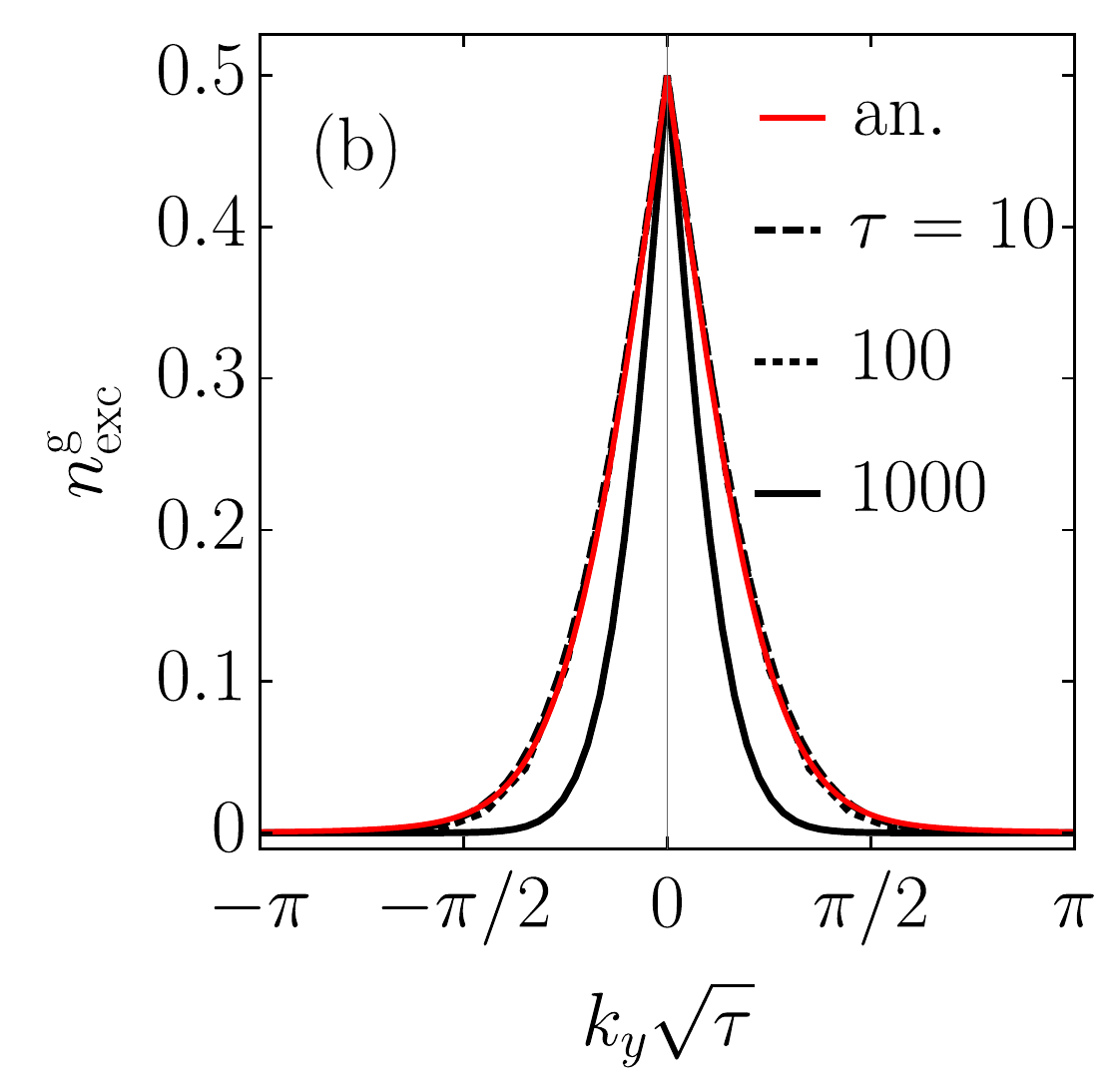}}
	
	\caption{ (a) Momentum distribution of bulk excitations in a ribbon with $N_x=70$ after quenches with $\tau=10$ (dashed), $\tau=20$ (dotted), and $\tau=50$ (full). Analytical result of Eq.~\eqref{eq:LZfull} is shown by the red line. (b) Momentum distribution of in-gap excitations in a ribbon with $N_x=20$ after quenches with $\tau=10$ (dashed), $\tau=100$ (dotted), and $\tau=1000$ (full). A rescaled ($\tau\rightarrow2.6\tau$) analytical result of Eq.~\eqref{eq:pExc} in Appendix~\ref{app:bandProp} is shown by the red line.}
	\label{fig:KExc}
\end{figure}

The topological transition at $u_c=-2$ is characterized by the fact that, in an extended system with periodic boundary conditions in both directions, the valence and the conduction bands form a Dirac cone at $\mathbf{k}=0$ at that value of $u$. In the ribbon geometry, the valence and the conduction bands split into a set of subbands. Additionally, on the topological side of the transition two in-gap bands appear. The (avoided) crossing of the in-gap bands is, due to the inversion symmetry and the chiral symmetry, pinned to $k_y=0$ and $\varepsilon=0$, respectively. In the ground state, single-electron states below the crossing are fully occupied while those above it are empty.

Therefore, when during a quench the parameter $u$ reaches this critical point, the energy gap both to conduction subbands as well as to excited in-gap states is minimal at $k_y=0$ and excitations occur predominantly around that point. The final occupancy of energy bands after the quench with $\tau=1$ is shown in Fig.~\ref{fig:BandExc}(a). As shown, the in-gap bands emerge and are partially populated. Excitations are also present in conduction and valence subbands representing the bulk of the ribbon and their number is maximal at $k_y=0$. This result has already been obtained in Refs.~\onlinecite{Caio2015,Duta17} where sudden quenches were performed. 

We now explore how the number of both kinds of excitations depends on the  quench time $\tau$. Let us first define the momentum distributions of the number of bulk and in-gap excitations,
\begin{equation}
\begin{split}
&n_\mathrm{exc}^\mathrm{b}(k_y)=\sum_{n=N_x+2}^{2N_x}\sum_{m=1}^{N_x}|\langle u_n(k_y)|\varphi_m(k_y)\rangle|^2,\\
&n_\mathrm{exc}^\mathrm{g}(k_y)=\sum_{n\in\{L,R\}}\sum_{m=1}^{N_x}|\langle u_{n}(k_y)|\varphi_m(k_y)\rangle|^2\Theta(\varepsilon_n(k_y)).
\end{split}
\end{equation}
Here $|u_n(k_y)\rangle$ is an eigenstate of the final Hamiltonian $\hat{H}(k_y)$ (for bulk excitations the sum over $n$ runs over the conduction subbands while for in-gap excitations it runs over the two in-gap bands), $\Theta$ is the Heaviside function, and $\{|\varphi_m(k_y)\rangle,m\in \{1,\ldots,N_x\}\}$ is the set of states obtained by time evolving the states that formed the valence band before the quench. 
 
Fig.~\ref{fig:TotExc} shows the number of excitations in the bulk $N_\mathrm{exc}^\mathrm{b}=\frac{1}{N_y}\sum_{k_y}n_\mathrm{exc}^\mathrm{b}(k_y)$  and in the in-gap states $N_\mathrm{exc}^\mathrm{g}=\frac{1}{N_y}\sum_{k_y}n_\mathrm{exc}^\mathrm{g}(k_y)$  as a function of quench time $\tau$. 
With increasing $\tau$ the number of bulk and edge excitations decreases. For a large range of $\tau$ from about $\tau\approx 1$ to the upper limit which increases with $N_x$, the number of in-gap excitation scales as $N_\mathrm{exc}^\mathrm{g}\propto \tau^{-1/2}$, while the number of excitations in the bulk scales as $N_\mathrm{exc}^\mathrm{b}\propto \tau^{-1}$.

To understand the origin of this difference it is instructive to first consider the momentum distribution of excitations. Results are shown for several $\tau$ in Fig.~\ref{fig:KExc} as a function of the scaled momentum $k_y\sqrt{\tau}$, where the left panel displays the distribution of the bulk and the right panel the distribution of the in-gap excitations. One immediately notices a crucial difference. The distribution of the bulk excitations and its dependence on the quench time can be essentially understood by integrating  standard Landau-Zener (LZ) result $n_\textrm{exc}^\textrm{pbc}(\mathbf{k})=\exp(-2\pi k^2/\alpha)$ with $\alpha=\pi(u_1-u_0)/\tau$, valid for a system with periodic boundary conditions in both directions,\cite{Unal16,Ulcakar2018} over $k_x$:
\begin{equation} n_\mathrm{exc}^\mathrm{pbc}(k_y)=\frac{N_x}{2\pi}\sqrt{\frac{\alpha}{2}}\exp(-2\pi k_y^2/\alpha). 
\label{eq:LZfull}
\end{equation}
One can understand the good agreement of this expression with numerical results (see Fig.~\ref{fig:KExc}(a)) from the fact that the edges do not matter importantly for the behavior of the bulk (some corrections become apparent for longer quench times only).
The momentum distribution of the in-gap excitations, in contrast, behaves differently. It can be described by a function which has a sharper peak close to $k_y=0$ and reaches value 1/2 there.  On the other hand, as presented on the plot, still different curves collapse when the momentum is scaled with $\sqrt{\tau}$ (with deviations at long quench times that we discuss later).

One can exploit the observed scaling of momentum distributions to explain the scaling of the number of excitations. The total number of excitations is given by $L^d/(2\pi)^d\int\mathrm{d}^d\mathbf{k}n_\textrm{exc}(\mathbf{k}) = \int\mathrm{d}^d\mathbf{k}g(k\sqrt{\tau}) \propto \tau^{-d/2}$, where $L^d$ is the volume of the system. Hence, for in-gap states, $N_\mathrm{exc}^\textrm{g} \propto \tau^{-1/2}$.  It is important that the integral was done in $d=1$. If the calculation was repeated for the bulk with periodic boundary conditions it is the $d=2$ that gives the scaling that holds for the bulk states (whereas for the case of a ribbon, the role of $k_x$ momentum that is not conserved due to the edges is played by the band-index). 

Now let us discuss how to rationalize the momentum distribution of edge-state excitations as documented by Fig.~\ref{fig:KExc}(b). We give a simplistic picture based on the evolution of eigenenergies  as displayed in Fig.~\ref{fig:BandExc}(b). One can find a continuous evolution of the lowest conduction state and the highest valence state into the edge states on crossing the critical point at $u=u_c$ from the non-topological side. A rough picture of the behavior is obtained by retaining these two states only. These approach $\varepsilon=0$ roughly linearly in $u-u_c$ on the non-topological side and roughly do not evolve any further on the topological side. This is a half of the two-band Landau-Zener problem for a system described by a low-energy two band Hamiltonian $\hat{H}_2(k_y)$, whose eigenenergies are shown in Fig.~\ref{fig:BandExc}(b) (red). The corresponding calculation is presented in Appendix~\ref{app:bandProp}. This predicts a function that matches closely the observed behavior, provided one additionally rescales the quench time by a constant. This constant is a reminder that the presented analysis can be at best considered only as a lowest order expansion of the multi-level Landau-Zener problem that retains only two levels, and the precise derivation of the behavior is left for future work.

As observed in Fig.~\ref{fig:TotExc}, power-law scalings of the number of excitations are valid only in a certain range of quench times. If a quench is performed too quickly, excitations are created also far away from the $k_y=0$, where the dynamics cannot be correctly described by the LZ model. This happens when $\tau\lesssim1$. If a quench is performed too slowly, the system evolves adiabatically across the critical point. The relevant energy gap here is the level spacing which scales as $N_x^{-1}$ near a critical point. As in the LZ physics the square of the energy gap determines the boundary between adiabatic and non-adiabatic evolution, the dynamics becomes adiabatic for $\tau\gtrsim N_x^2$. The number of bulk excitations drops exponentially with quench time in this regime. On the other hand, the dynamics of in-gap states is non-adiabatic even for very long quench times, but only deep in the topological regime where our approximation with two diabatic levels completely fails.  Therefore, the $\tau^{-1/2}$ scaling also holds only for $\tau\ll N_x^2$.

At last, we discuss whether the QWZ model follows the KZ scaling, which predicts that for linear quenches the total number of excitations scales with the quench rate as $N_\mathrm{exc}\propto\tau^{-d\nu/(\nu z+1)}$. Here,  $\nu$ and $z$ are the correlation length and the dynamical critical exponent, respectively, associated with the critical point across which the system is quenched. Refs.~\onlinecite{Damski05,Dutta10,Ulcakar2018} considered two-dimensional systems with periodic boundary conditions that form a Dirac cone at the critical point and showed that according to the KZ mechanism the number of bulk excitations scales as $\tau^{-1}$. To perform the corresponding calculation for in-gap states, we again described those approximately with the Hamiltonian $\hat H_2(k_y)$. Following Ref.~\onlinecite{Chen16} (see Appendix~\ref{app:critExp}), we obtained the critical exponents for the in-gap states. These read $\nu=1$ and $z=1$ and, taking into account the one-dimensional nature of in-gap states, correctly reproduce the scaling of in-gap excitations as found using the Landau-Zener approach. 

\subsection{Hall conductance}
\label{sec:HallCond} 
\begin{figure}[h]
	\centering
	{\includegraphics[width=125pt]{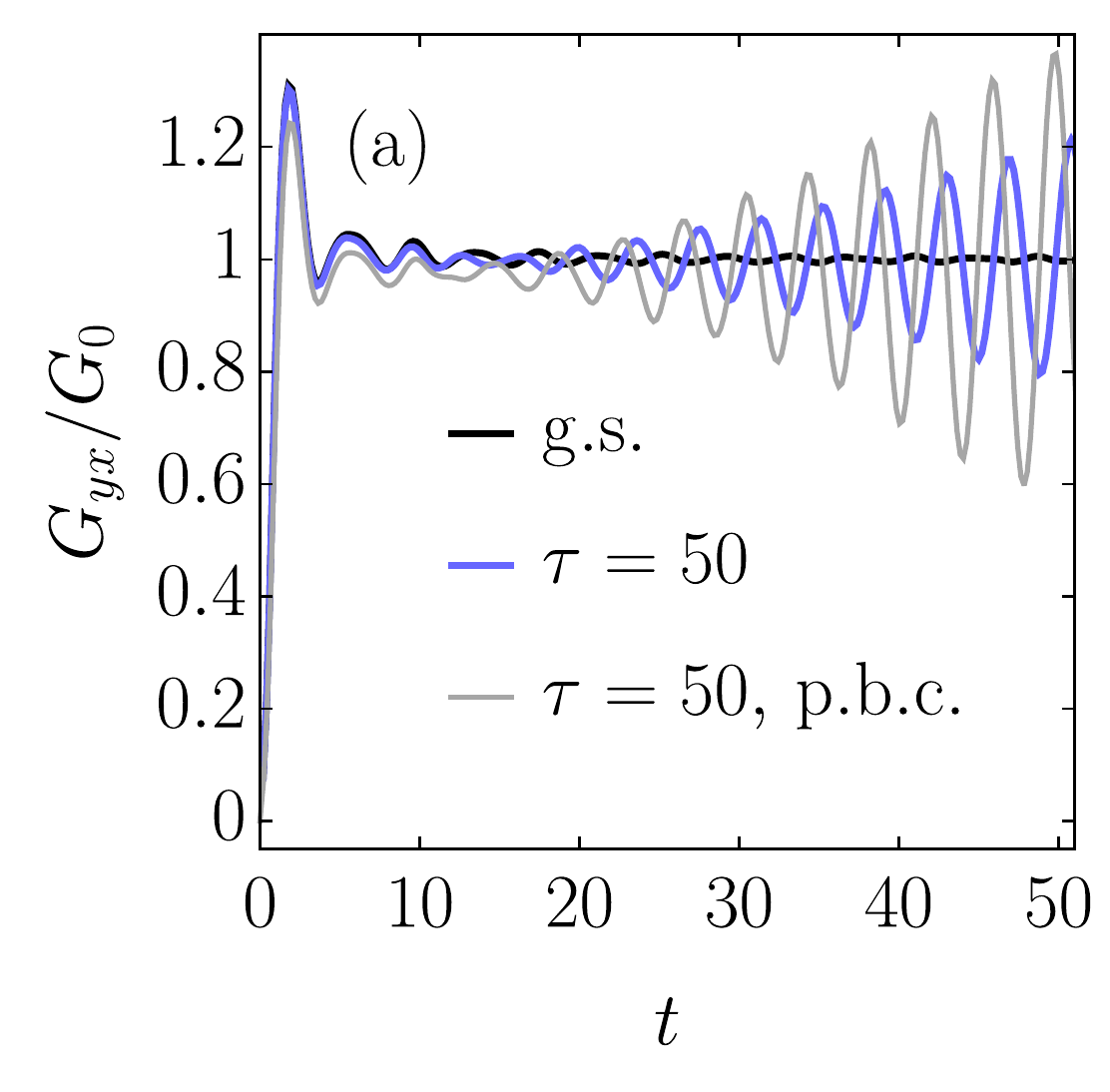}\includegraphics[width=125pt]{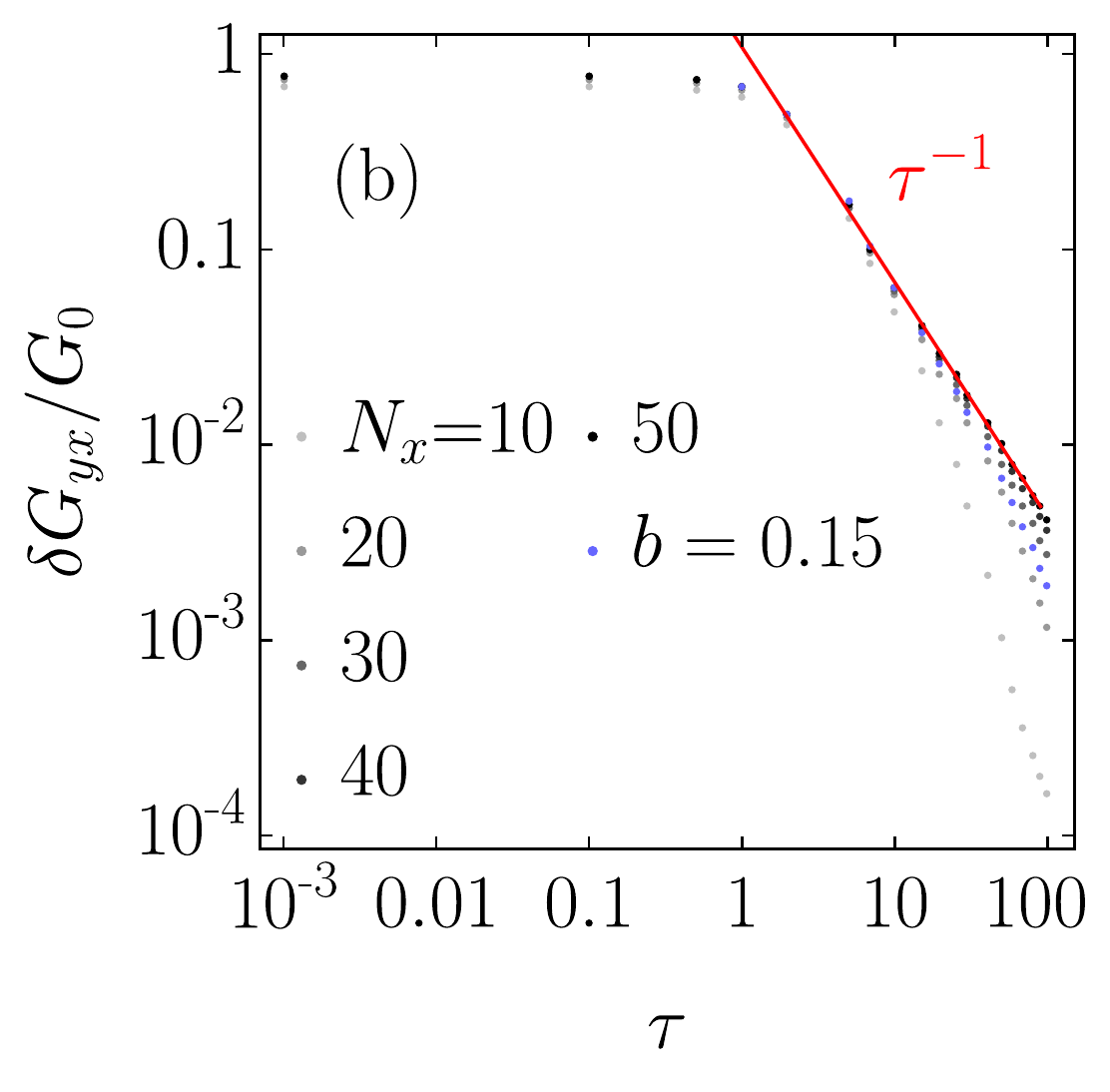}}
	\caption{ a) Hall conductance of the ground state of the final Hamiltonian (black) and after a quench with $\tau=50$ (blue), and Hall conductivity after a quench with $\tau=50$ in a system with periodic boundary conditions in both spatial directions (gray), for $N_x=20$, $E_0=0.001$ and $\tau_E=5$. (b) Deviations from the ground state Hall conductance as a function of $\tau$ for different ribbon widths, ranging from $N_x=10$ (light gray) to $N_x=50$ (black). The red line is a fitted $\tau^{-1}$ scaling. Blue dots show results for a Hamiltonian with a broken inversion symmetry of magnitude $b=0.15$ and $N_x=20$.}
	\label{fig:QSxy}
\end{figure}
We now turn to the Hall response. In Fig.~\ref{fig:QSxy}(a) we show the Hall conductances of two states of a ribbon, namely of its ground state at $u=u_1$ and of the state created by a slow quench, together with the Hall conductivity of a system with periodic boundary conditions in both directions, again after a slow quench. In all of the cases the response first experiences a transient behavior and then oscillates around a non-zero value $\bar{G}_{yx}$, with the frequency corresponding to the band gap of the final Hamiltonian. The latter case represents a pure bulk response which was already studied in Refs.~\onlinecite{zoller,Unal16,Caio2016,Ulcakar2018}. In Ref.~\onlinecite{Ulcakar2018} it was shown that the long-time average of the Hall conductivity approaches the ground-state value with the deviations scaling with quench time as $\tau^{-1}$. Deviations of $\bar{G}_{yx}$ of a quenched ribbon from its quantized ground state value $G_0=q^2/h$, $\delta G_{yx}=|\bar G_{yx}-G_0|$, are presented in Fig.~\ref{fig:QSxy}(b). For quenches faster than $\tau\approx1$ the average Hall conductance almost vanishes. For slower quenches the average Hall conductance increases towards the quantized value. In wide ribbons, the deviations diminish as $\tau^{-1}$, as in systems with periodic boundary conditions in both directions.

One might expect the excitations in the in-gap states to dominate the deviations and hence anticipate a scaling with their number that goes as $\tau^{-1/2}$, instead. In order to understand the observed behavior, we evaluated the Hall conductance using the time-dependent perturbation theory. One obtains the following result for the time-averaged Hall current at large times (see Appendix~\ref{app:HallDer}):
\begin{equation}
\bar{I}_{y}=G_0E_0\sum_{n=1}^{2\,N_x}\int\mathrm{d}k_y\,n_n(k_y)\Omega_{n}(k_y),\label{eq:SxyAver}
\end{equation}
\begin{equation}
\Omega_{n}(k_y)=-\partial_{k_y}\langle u_n(k_y)|\hat{x}|u_n(k_y)\rangle,\label{eq:BerryCurv}
\end{equation}
where $n_n(k_y)$ is the occupation of the eigenstate $|u_n(k_y)\rangle$ of the final Hamiltonian. The Hall conductance is then calculated as $\bar G_{yx}=\bar I_y/U_x$ where $U_x=E_0(\langle u_R(0)|\hat x|u_R(0)\rangle-\langle u_L(0)|\hat x|u_L(0)\rangle)$ is the voltage between the edges of the ribbon (see Appendix~\ref{app:GSelectric}). The Hall conductance is expressed as an integral of the quantity $\Omega_n(k_y)$ weighted by the band occupancy. This result comes from the fact that the current is given by the change of the dispersion due to the electric potential $-qE_x\hat x$, which leads to terms of the form $\partial_{k_y} \langle \hat{x} \rangle$. If the system is in the ground state, $n_n(k_y)$ is equal to the Fermi distribution $f(\varepsilon_n(k_y))$ and the conductance is quantized as shown in Eq.~\eqref{eq:IasBcurvGS} in Appendix~\ref{app:GSelectric}. The deviation of the post-quench Hall conductance from the ground state one can thus be attributed to excitations: $\delta G_{yx}\propto\sum_{n=1}^{2N_x}\int\mathrm{d}k_y\,\delta n_n(k_y)\Omega_{n}(k_y)$ where $\delta n_n(k_y)=n_n(k_y)-f(\varepsilon_n(k_y))$. 

Let us first analyse the in-gap contributions to $\delta G_{yx}$. In the topological phase, quantities $\Omega_{L,R}(k_y)$ are linear in $k_y$ in the interval around $k_y=0$ where excitations are present after slow quenches (see Appendix~\ref{app:HallDer1}). As $\delta n_{L,R}(k_y)=\pm n_\textrm{exc}^\textrm{g}(k_y)\mathrm{sgn}(k_y)$, the  in-gap contribution to $\delta G_{yx}$ is proportional to $\int\mathrm{d}k_y\,n_\mathrm{exc}^\mathrm{g}(k_y)|k_y|=\int\mathrm{d}k_y\,g(k_y\sqrt{\tau})|k_y|\propto \tau^{-1}$. This is the same scaling as for the "bulk-only" system with periodic boundary conditions in both directions. 

On the other hand, $\Omega_n(k_y)$ for bulk subbands take a finite value at $k_y=0$. After slow quenches, subbands near the bottom of the conduction band and those near the top of the valence band carry most of the bulk excitations. Therefore we can, to a good approximation,  evaluate $\delta G_{yx}$ by replacing $\Omega_n(k_y)$ for conduction and valence subbands in Eq.~\eqref{eq:SxyAver} with values at $k_y=0$ for the lowest conduction and the highest valence subband, respectively. As these two values differ only in sign, this approximation leads to a deviation of the post-quench time-averaged Hall conductance from the ground state value which scales as the number of bulk excitations, i.e. as $\tau^{-1}$, as is also the case in a system with periodic boundary conditions in both directions.   

\subsection{Breaking of the inversion symmetry}\label{sec:breaking}

As shown in Sec.~\ref{sec:model} the QWZ model in the ribbon geometry possesses the inversion symmetry which pins the in-gap subbands crossing to $k_y=0$ during quenches performed in this work. If the inversion symmetry is broken, as is, e.g., the case in the Haldane model studied in Ref.~\onlinecite{Privitera16}, the $k_y$-point where the crossing occurs may move during the quench. After the quench, this results in a perfect population of the excited in-gap states in the interval of $k_y$-s swept over by the crossing. As discussed in Ref.~\onlinecite{Privitera16}, the reason behind this behavior is that, at each of these momenta, an electron performs a complete Landau-Zener tunneling process with unit transition probability due to an exponentially small Landau-Zener gap.

To break the inversion symmetry in the QWZ model, we add the term $\sum_{x=1}^{N_x-1}|x+1\rangle\langle x|\otimes b\frac{-i \hat{\sigma}_x-\hat{\sigma}_y}{4}+\mathrm{h.c.}$ to the Hamiltonian $\hat{H}(k_y)$. As the parameter $u$ enters the topological regime at $u_c=-2$, the in-gap subbands exhibit the crossing at a $k_y$ of the order of $b$. As the quench progresses, the crossing shifts towards $k_y=0$. Therefore, the in-gap excitations are generated in an interval of $k_y$ of width of the order of $b$. On the other hand, if the inversion symmetry is not broken, in-gap excitations are generated in an interval of $k_y$ of width of the order of $\tau^{-1/2}$. For slow quenches, $\tau\gg b^{-2}$, the number of in-gap excitation thus saturates while in the opposite limit, $\tau\ll b^{-2}$, the effect of the inversion symmetry breaking on the scaling of the in-gap excitations on $\tau$ can be neglected. This can be seen in Fig.~\ref{fig:GapExc}(b) where blue dots show the scaling of the number of in-gap excitations with quench time for $b=0.15$. In contrast, the scaling of the bulk excitations, shown in Fig.~\ref{fig:GapExc}(a), is not affected by the inversion-symmetry breaking.

The mechanism described above leads to asymmetrically populated in-gap bands after a quench with $b\ne0$. This results in a finite current in $y$-direction already in the absence of electric field. This is not observed in the presence of the inversion symmetry as there a quench by itself cannot lead to a finite current as excitations are produced symmetrically in in-gap bands, $n_L(k_y)=n_R(-k_y)$, which, taking into account the fact that the corresponding dispersion curves have opposite slopes, leads to a perfect cancellation of current. Turning on the electric field in $x$-direction induces an additional Hall current, which, as shown with blue dots in Fig.~\ref{fig:QSxy}(b), is not significantly affected by the inversion-symmetry breaking.

We note that not every perturbation breaking the inversion symmetry leads to a violation of the KZ scaling. A perturbation may break the symmetry of the energy spectrum and shift the in-gap crossing to $k_y\neq0$, however as long as the in-gap crossing is pinned to a certain $k_y$ throughout the quench, the KZ scaling holds. The inversion symmetry assures KZ scaling but is not necessarily needed.

\subsection{Breaking of the chiral symmetry}\label{sec:breakchiral}

In Sec.~\ref{sec:model} we showed that a QWZ ribbon possesses also the chiral symmetry due to which the in-gap bands crossing is pinned to zero energy. Here we break this symmetry by adding a term $\sum_{x=1}^{N_x-1}|x+1\rangle\langle x|\otimes b\hat1+\mathrm{h.c.}$ to $\hat{H}(k_y)$. During the quench, the in-gap crossing moves along the $k_y=0$ line (due to  the inversion symmetry), starting at $\varepsilon\propto b$ at $u=-2$ and moving towards zero energy at $u=-1$. We performed quenches for several values of $b$ up to $b=0.3$. Breaking the chiral symmetry in this way did not affect the scaling of the number of excitations, neither the scaling of deviations of the Hall conductance. The KZ scaling still holds because the momentum distribution of excitations is not affected and still behaves as $n_\mathrm{exc}(\mathbf{k})=g(k\sqrt{\tau})$. This can be seen by considering the in-gap two-level Landau-Zener problem (see Eq.~\eqref{eq:H2}) with additional term which shifts the in-gap crossing from zero energy: the minimal energy gap during the quench remains proportional to $|k_y|$ and the velocity $\alpha$ with which the two energy bands approach is unchanged.

\section{Conclusions}
\label{sec:concl}

To conclude, we investigated the nonequilibrium state that arises in Chern topological insulator ribbons after a slow quench from the trivial to a topological regime. During such a quench excitations are generated both in the bulk bands as well as in the in-gap states localized at the edges of a ribbon. The number of excitations in the bulk drops with quench time $\tau$ as $\tau^{-1}$ while the number of in-gap excitations, provided the ribbon possesses the inversion symmetry, drops as $\tau^{-1/2}$. While the former result is well known from the studies of bulk systems, we explained the latter by taking into account that the in-gap bands, once they form upon entering the topological regime, are stationary. This allowed us to approximately map the quench dynamics related to the in-gap bands to a half of the Landau-Zener problem. Taking additionally into account the one-dimensional nature of the in-gap states we derived the $\tau^{-1/2}$ scaling. We also calculated the critical exponents of the topological transition and showed that, in contrast to the behavior of systems where the inversion symmetry is broken, the scaling of in-gap excitations follows the Kibble-Zurek prediction. Finally, we characterized the quenched state by calculating the Hall conductance. We found that its deviations from the quantized value characteristic of the post-quench Hamiltonian drop as $\tau^{-1}$, which was also found in systems with periodic boundary conditions.

\acknowledgements

We acknowledge useful discussions with A. Ram\v{s}ak, J. Rozman, and D. Monaco. The work was supported by the Slovenian Research Agency under contract no. P1-0044.

\appendix

\section{Ground state response to the electric field}

\label{app:GSelectric}
\begin{figure*}
	\centering
	{\includegraphics[width=125pt]{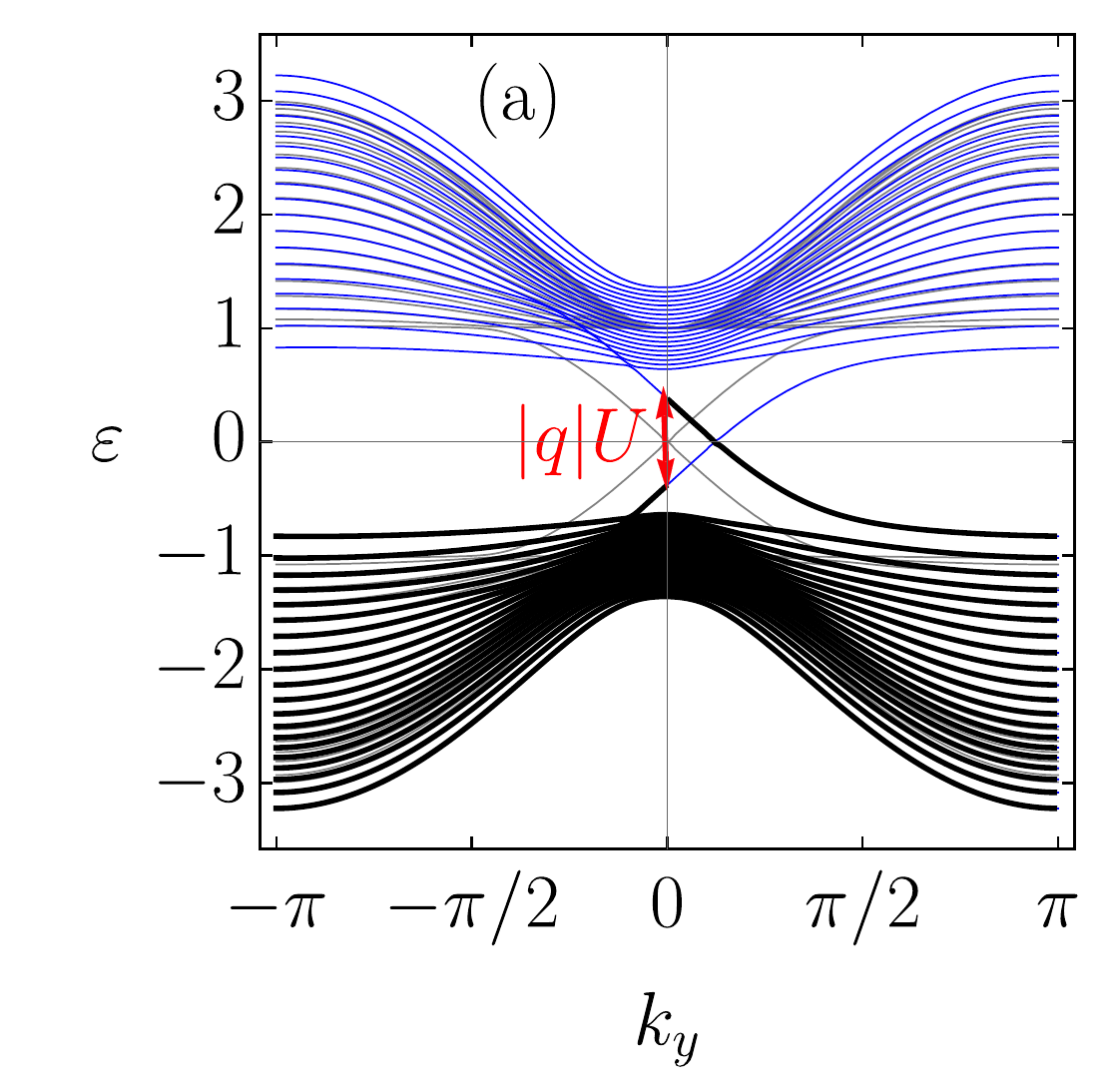}\includegraphics[width=125pt]{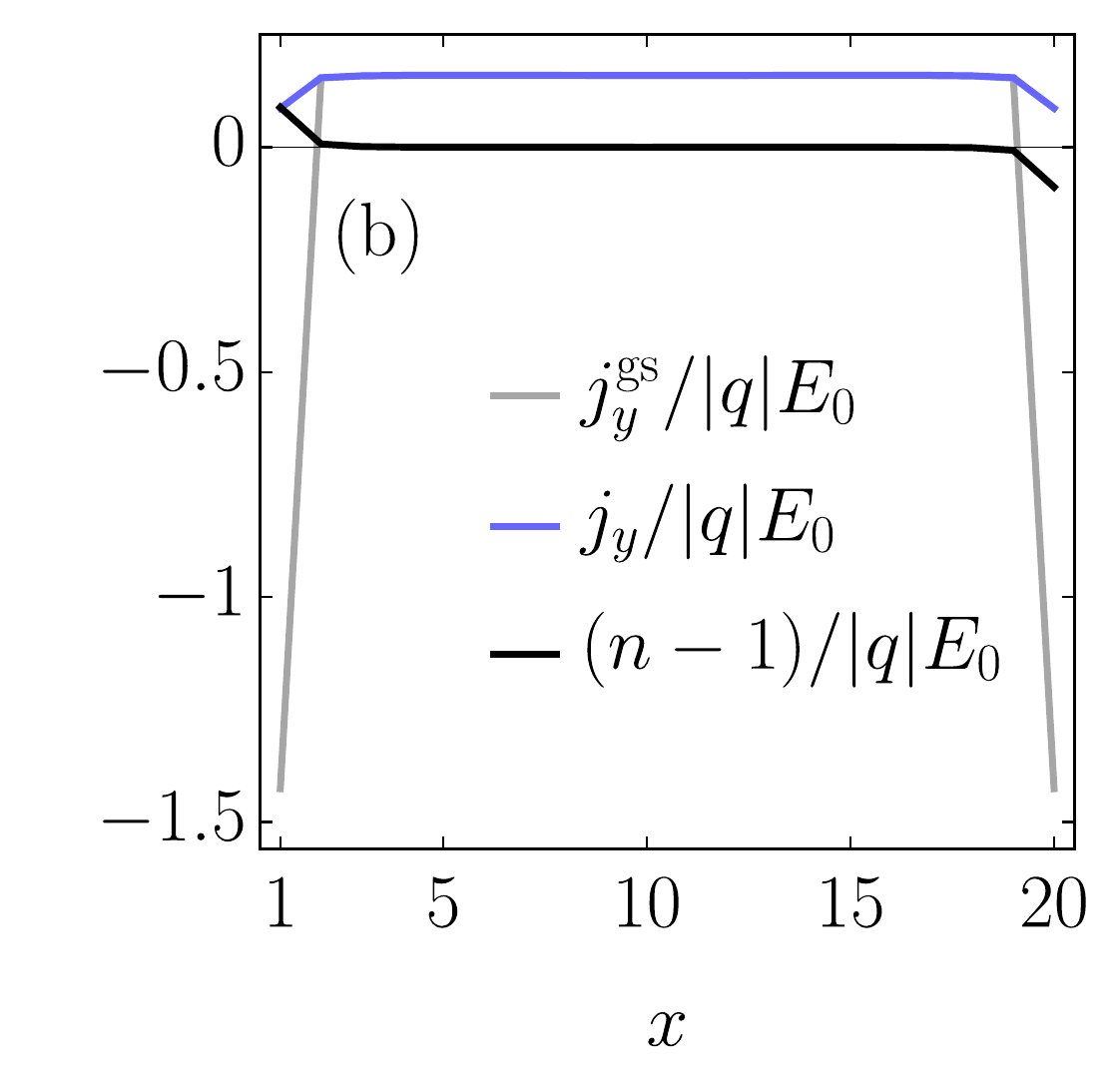}\includegraphics[width=125pt]{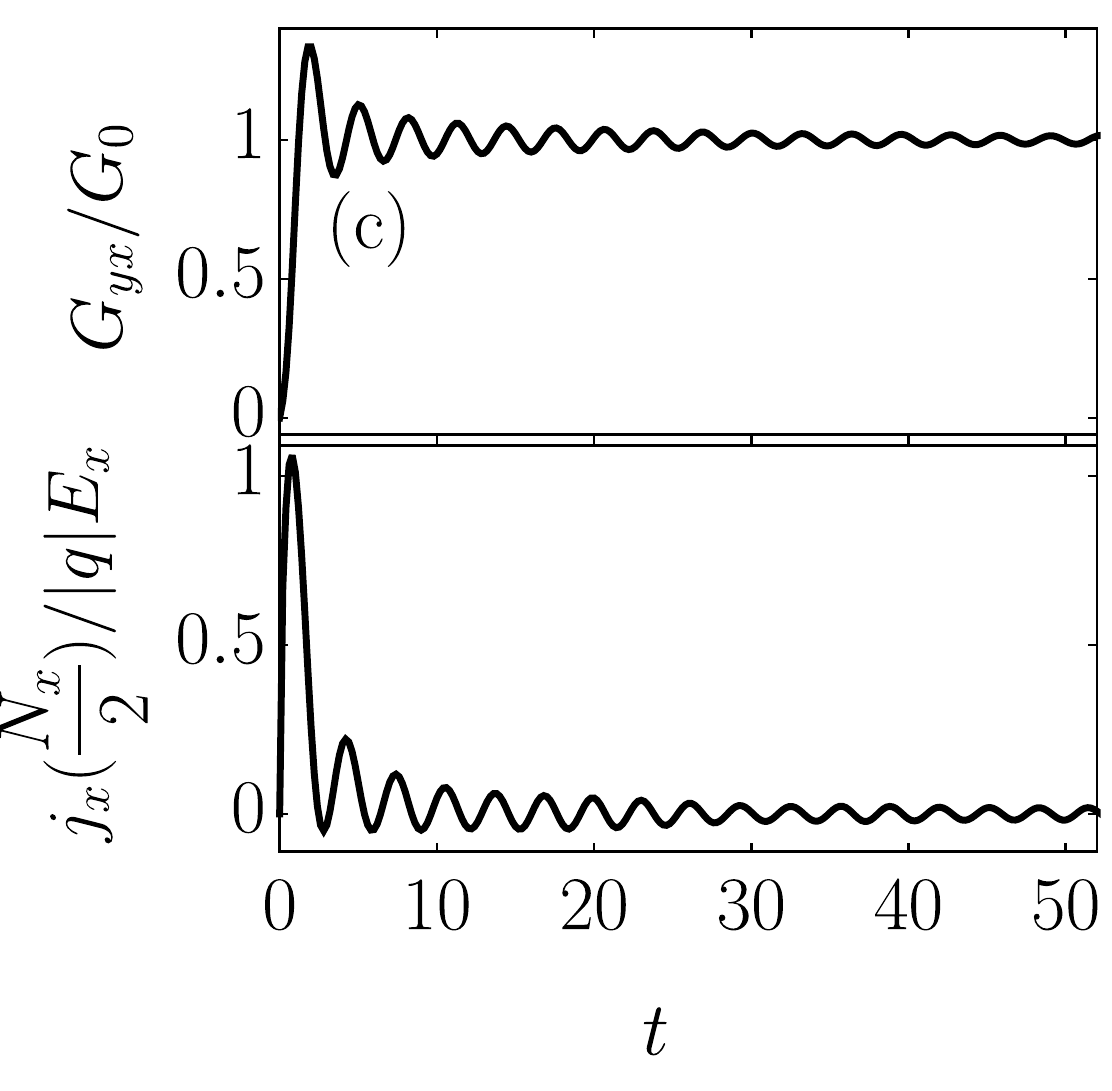}}
	\caption{a) Energy spectrum in the topologically non-trivial phase in absence (gray) and in presence of the electric field (blue and black). Black lines denote the occupied states of the system that adiabatically evolved due to insertion of the electric potential from the ground state without the electric potential. b) Current densities $j_y^\mathrm{gs}$ in the ground state of the system with a static electric field (gray) and $j_y$ in the state which adiabatically evolved from the ground state of the system without the electric field at $t\gg \tau_E$ (blue). 
	The deviation of the particle density $n-1$ is also shown (black). c) Conductance $G_{yx}$ and the current density in $x$-direction in the middle of the ribbon $j_x(N_x/2)$ as a response to the adiabatically inserted electric potential. The system size is $N_x=20$, $N_y=201$, $u=-1$ and the electric field $E_0=0.04$, $\tau_E=5$.}
	\label{fig:GS}
\end{figure*}

In the ground state in absence of the electric field the current density $\langle\hat{\boldsymbol{j}}(\mathbf{r})\rangle=(j_x(\mathbf{r}),j_y(\mathbf{r}))$ is zero throughout the system and the particle density $n(\mathbf{r})=\sum_{k_y}\sum_{n=1}^{N_x}|\langle\mathbf{r}|\Psi_n(k_y)\rangle|^2$ is equal to $1$ on every site.
As the electric field is turned on adiabatically in the topologically non-trivial phase, the current and the particle densities develop profiles shown in Fig.~\ref{fig:GS}(b). The current starts flowing in $y$-direction mostly through the bulk, while the density deviates from 1 at the edges: on the $x=1$ edge where the electric potential is higher the particle density is increased while on the opposite edge where the electric potential is lower the particle density is lowered. Note that these profiles are invariant of the electric field strength $E_0$ as long as it can be treated as an adiabatic perturbation. Also, the current density in the bulk is constant and invariant of $N_x$. 

The time-dependence of the Hall conductance is shown in Fig.~\ref{fig:GS}(c). The conductance rises steeply and then oscillates around a quantized value of $1$ in units of $G_0=q^2/h$ with the frequency corresponding to the bulk energy gap. The amplitude of the oscillations diminishes with time and it also becomes smaller if the electric field is turned on more adiabatically (i.e., with a longer $\tau_E$).
In contrast to the Hall conductance, the current density $j_x$ oscillates around zero except for small times $t\lesssim \tau_E$ where it is positive. This corresponds to the current flowing from $x=1$ to $x=N_x$ until the final inhomogeneous particle density profile is established. 

For completeness let us also consider the case when the electric field is adiabatically turned on in the topologically trivial phase. The particle density and the current density $j_x$ behave as in the topologically non-trivial phase discussed above. The current density $j_y$ is non-vanishing, however the time average of the total current $I_y$, proportional to the Hall conductance, is exactly zero.

The quantization of the Hall conductance can be explained by comparing the adiabatically evolved system to: (1) the system in the ground state in presence of the electric field; (2) the system in the ground state without electric field.

\subsection{Comparison to the ground state with electric field} 

In presence of the electric field, the energy bands are deformed as shown in Fig.~\ref{fig:GS}(a) (blue and black lines). According to the first order perturbation theory, the eigenenergy of an eigenstate $|\Psi_n(k_y)\rangle$ changes due to the perturbing potential $\hat{V}$,
\begin{equation}
\begin{split}
\delta\varepsilon_n(k_y)=\langle \Psi_n(k_y)|\hat{V}|\Psi_n(k_y)\rangle=\\
-qE_x\langle u_n(k_y)| \hat{x}-\tfrac{N_x+1}{2}|u_n(k_y)\rangle.
\label{firstorder}
\end{split}
\end{equation}
As the in-gap states near $k_y=0$ are localized at opposite edges, their eigenenergies change significantly. The change of the energy dispersion results in a shift of the degeneracy point of the in-gap states from $k_y=0$ to $k_y=k_0>0$.

In the ground state of the system with static electric field the total current is zero. This can be explicitly seen by summing all the contributions to the current carried by the occupied eigenstates, where the contribution of an eigenstate  $|\Psi_n(k_y)\rangle$ is equal to \cite{madzari}
\begin{equation}
\langle \Psi_n(k_y)|\hat{I}_y|\Psi_n(k_y)\rangle=\frac{q}{N_y}\partial_{k_y}\varepsilon_n(k_y).
\label{eq:IyEigen}
\end{equation}
The calculated current density $j_y^{\mathrm{gs}}(\mathbf{r})$ is shown in Fig.~\ref{fig:GS}(b) (gray). As the current density in the bulk is constant, the edge current flows in the opposite direction so it completely cancels out the bulk contributions to the net current.

The quantization of the Hall conductance can be explained by comparing the occupations of the energy bands in the adiabatically-evolved system with those in the system in the ground state with static electric field. As the electric field is turned on, the electrons adiabatically follow the bands which get deformed. This results in the final occupancy, shown in Fig.~\ref{fig:GS}(a) (black lines), which is different from that in the ground state of the Hamiltonian with electric field. The difference lies only in the in-gap bands: band $L$ is occupied up until $\varepsilon_L(0) = -qE_x\langle u_L(0)| \hat{x}-\frac{N_x+1}{2}|u_L(0)\rangle$, while band $R$ is occupied up until $\varepsilon_R(0)=-qE_x\langle u_R(0)| \hat{x}-\frac{N_x+1}{2}|u_R(0)\rangle$. The same occupancy would be observed in a quantum wire contacted to two electron reservoirs, one having the chemical potential at $-qU/2$ and the other at $qU/2$ with $U=E_x(\langle u_R(0)| \hat{x}|u_R(0)\rangle-\langle u_L(0)|\hat{x}|u_L(0)\rangle)$.
As the total current in the ground state is zero, only the difference in occupations with respect to those in the ground state contributes to the current,
\begin{equation}
\delta I_y=\frac{q}{2\pi}\left(\int_{\varepsilon_{R}(0)}^{\varepsilon_{R}(k_0)}\mathrm{d}\varepsilon_{R}-\int_{\varepsilon_{L}(0)}^{\varepsilon_{L}(k_0)}\mathrm{d}\varepsilon_{L}\right)=G_0U.
\end{equation}
$U$ is equal to the voltage between ribbon edges measured by attaching voltage probes as the chemical potential at $x=1$ edge is equal to $\varepsilon_L(0)$ while the chemical potential at the $x=N_x$ edge is equal to $\varepsilon_R(0)$. Therefore, the Hall conductance is quantized in units of $G_0$.

\subsection{Comparison to the ground state without electric field}
\label{sec:GScurBerry}
The quantization of the Hall conductance in a system with adiabatically inserted electric field can be also explained by comparing it to the system in the ground state without electric field. As the electrons in the adiabatically evolved system are occupying eigenstates with perturbed eigenenergies, the additional current due to the electric field carried by an electron in an eigenstate $|\Psi_n(k_y)\rangle$ is according to Eqs.~\eqref{firstorder} and \eqref{eq:IyEigen} equal to $\frac{q^2E_x}{N_y}\Omega_n(k_y)$, where we denoted $\Omega_n(k_y)=-\partial_{k_y}\langle u_n(k_y)|\hat{x}|u_n(k_y\rangle$. As the total current in the ground state is zero, only this contribution needs to be taken into account, resulting in
\begin{equation}
\begin{split}
&\delta I_y=\frac{q^2E_x}{2\pi}\sum_{n=1}^{2N_x}\int_{-\pi}^{\pi}\mathrm{d}k_y 
f(\varepsilon_n(k_y)) \Omega_n(k_y)=\\
&G_0E_x\Big(-\langle u_L(k_y)| 
\hat{x}|u_L(k_y)\rangle\Big|_{-\pi}^{0}-\\
&\langle u_R(k_y)| \hat{x}|u_R(k_y)\rangle\Big|_{0}^{\pi}\Big)=G_0U,
\label{eq:IasBcurvGS}
\end{split}
\end{equation}
where $f(\varepsilon)$ is the Fermi distribution.

\section{Number of in-gap excitations}
\label{app:bandProp}

In this section we derive the number of in-gap excitations for quenches with $1\ll\tau\ll N_x^2$, which is the regime where the power-law scaling of the number of excitations holds. In this regime, the time-evolution of the system is strongly non-adiabatic at the critical point. On the other hand, quenches are slow enough for the excitations to be formed only close to the $k_y=0$ point where the LZ calculation applies.

\begin{figure*}
	\centering
	\includegraphics[width=450pt]{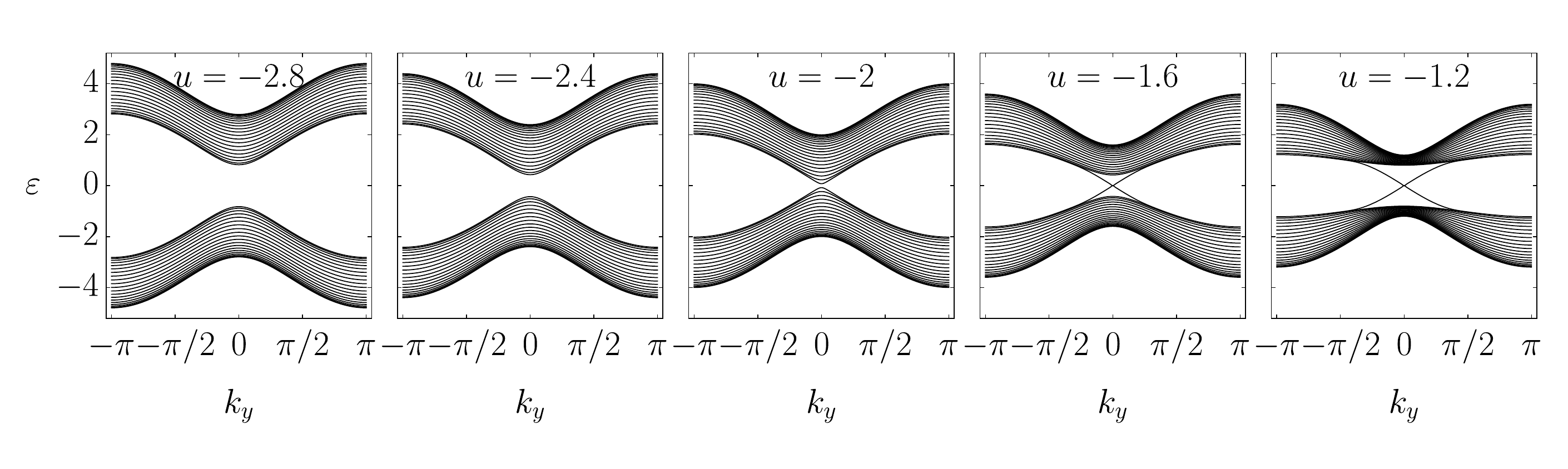}
	\caption{Energy dispersion at different values of parameter $u$ during the quench.}
	\label{fig:bandsU}
\end{figure*}

\begin{figure}[h]
	\centering
	{\includegraphics[width=125pt]{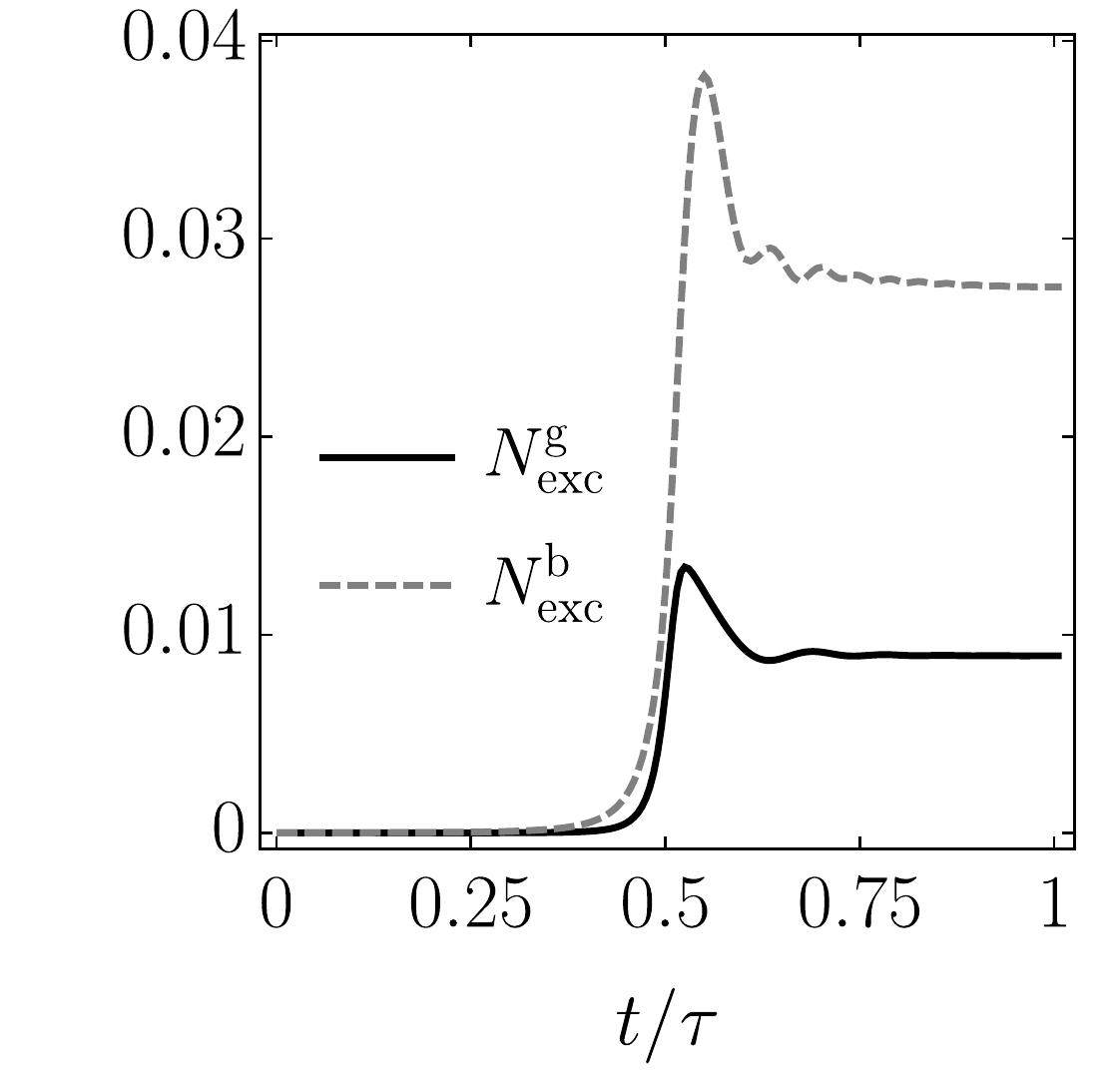}}
	\caption{Total number of excitations in the in-gap states (solid) and in the bulk (dashed) during the quench with $\tau=80$ and $N_x=50$.}
	\label{fig:GapExc}
\end{figure}

The energy spectra at different values of parameter $u$ are shown in Fig.~\ref{fig:bandsU}.
While the system is in the trivial phase, there are no in-gap states present and all of the states are delocalized  throughout the whole sample. Near the critical point at $u_c=-2$ the in-gap bands are formed. As $u$ enters the topological phase, the energy gap between the bulk bands reopens while the dispersion of the in-gap bands appears to be static for $k_y$ close to $0$. Fig.~\ref{fig:GapExc} shows the total number of in-gap excitations during the quench with $\tau=80$ and $N_x=50$. The total number of excitations jumps to a non-zero value when the system is near the critical point at $t/\tau=1/2$. From the critical point on, the total number of excitations is approximately constant. In our analytical evaluation we will therefore assume that the number of in-gap excitations does not change from the critical point on.

In order to calculate the number of excitations in the in-gap states, we first express the Hamiltonian $\hat{H}(k_y)$ in the diabatic basis $\{|\psi_{n\sigma}\rangle=\sum_{x=1}^{N_x}\sqrt{\frac{2}{N_x+1}}\sin(k_nx)|x\rangle\otimes|\sigma\rangle,\,k_n=\frac{\pi n}{N_x+1},\,n\in\{1,\ldots N_x\},\,\sigma\in\{A,B\}\}$. $|\psi_{n\sigma}\rangle$ is an eigenstate of the Hamiltonian that does not couple the $|A\rangle$ and $|B\rangle$ orbitals (when $\hat{\sigma}_x$ and $\hat{\sigma}_y$ terms are put to zero in Eq.~\eqref{eq:H}) with the eigenenergy $\varepsilon_{n\sigma}(k_y)=\pm(u+\cos k_y+\cos k_n)$. Near the critical point $u$ and therefore eigenenergies of the diabatic states vary linearly in time, which corresponds to the multi-level LZ paradigm. As this problem is not exactly solvable, we approximate the dynamics of electrons in the in-gap bands with a two-level Hamiltonian
\begin{equation}
\hat{H}_2(k_y)=(u+\cos k_y+\cos k_1)\hat{\sigma}_z+\sin k_y\hat{\sigma}_y,\label{eq:H2}
\end{equation}
which is the projection of the full Hamiltonian~\eqref{eq:H} to the subspace spanned by diabatic states $|\psi_{1A}\rangle$ and $|\psi_{1B}\rangle$. The energy levels of $\hat{H}_2(k_y)$ as functions of $u$ are shown with red color in Fig.~\ref{fig:BandExc}. $\hat{H}_2(k_y)$ recreates the exact eigenenergies well up until the critical point. Since the number of excitations is approximately constant from the critical point on, we evaluate the number of excitations from the LZ formula at $t=0$ (from here on we measure time from the critical point on). As the excitations are for slow quenches generated only at momenta $k_y$ close to $0$ and at times near the critical point, we approximate the Hamiltonian~\eqref{eq:H2} up to the first order in $t$ and $k_y$.  The dynamics is dictated by the time-dependent Schr\"{o}dinger equation,
\begin{equation}
i\frac{\mathrm{d}}{\mathrm{d}t}|\varphi(k_y,t)\rangle=\left[\frac{\alpha}{2}t\hat{\sigma}_z+k_y\hat{\sigma}_y\right]|\varphi(k_y,t)\rangle,
\label{eq:HbulkLZ}
\end{equation} 
where $\alpha=(u_1-u_0)\pi/\tau$. We write the wave function as 
\begin{equation}
|\varphi(k_y,t)\rangle=a(k_y,t)e^{\frac{i}{2}\int\alpha \mathrm{d}t}|\psi_{1A}\rangle+b(k_y,t)e^{-\frac{i}{2}\int\alpha \mathrm{d}t}|\psi_{1B}\rangle,
\end{equation}
with initial conditions $|a(k_y,-\infty)|=1$ and $b(k_y,-\infty)=0$. To obtain the probability for transition to an excited state when the energy gap is closed at $t=0$, we use the result by Zener,\cite{ZenerLandau}
\begin{equation}
\begin{split}
a(k_y,t)=e^{-i\alpha t^2/4}e^{-\pi\gamma/4}e^{-i\pi/4}\frac{|k_y|}{ik_y}\times\\
\left(D'_{-n-1}(iz)+\frac{z}{2}D_{-n-1}(iz)\right),\\
b(k_y,t)=e^{i\alpha t^2/4}\sqrt{\gamma}e^{-\pi\gamma/4}D_{-n-1}(iz),
\end{split}
\end{equation}
where $z=\sqrt{i\alpha}t$, $\gamma=k_y^2/|\alpha|$, $n=-ik_y^2/\alpha$ and $D_{n}$ is the Weber function.\cite{ModAn} The eigenstates of the Hamiltonian at $t=0$ are $|\phi_+\rangle=\frac{1}{\sqrt{2}}(-i\frac{k_y}{|k_y|}|\psi_{1A}\rangle+|\psi_{1B}\rangle)$ and $|\phi_-\rangle=\frac{1}{\sqrt{2}}(i\frac{k_y}{|k_y|}|\psi_{1A}\rangle+|\psi_{1B}\rangle)$ and the probability of finding an electron with $k_y$ in the state with the higher energy $|\phi_+\rangle$ is $n_\mathrm{exc}^\mathrm{g}(k_y)=\frac{1}{2}|i\frac{k_y}{|k_y|}a(k_y,0)+b(k_y,0)|^2$, its explicit expression being
\begin{equation}
n_\mathrm{exc}^\mathrm{g}(k_y)=\frac{\pi}{4}e^{-\frac{\pi k_y^2}{2\alpha}}\left|\frac{-1+i}{\Gamma\left(\frac{1}{2}+i\frac{k_y^2}{2\alpha}\right)}+\frac{|k_y|}{\sqrt{\alpha}\Gamma\left(1+i\frac{k_y^2}{2\alpha}\right)}\right|^2.
\label{eq:pExc}
\end{equation}
This analytical result only roughly describes the number of excitations, its deviations originating from the approximation that the in-gap states evolve as a two-level system. We fitted the function in Eq.~\eqref{eq:pExc} to the numerical results and obtained the parameter $\alpha=(u_1-u_0)\pi/(2.6\tau)$ that fits the actual number of excitations the best. The fitted function is shown in Fig.~\ref{fig:KExc}(b) (red curve). As the agreement between the fitted function and the actual number of excitations is perfect for small $k_y$, we are confident that the analytical result gives the correct scaling of the total number of in-gap excitations with quench time.

\section{Calculation of the critical exponents}
\label{app:critExp}
Let us choose a control parameter $\varepsilon$ such that the system undergoes a quantum phase transition at $\varepsilon=0$. A quantum phase transition is characterized by the divergence of both the characteristic length scale $\xi(\varepsilon)\propto\left|\varepsilon\right|^{-\nu}$ and the characteristic time scale $\tau(\varepsilon)\propto \left|\varepsilon\right|^{-z\nu}$, $\nu$ and $z$ being the correlation length and the dynamical critical exponents, respectively. According to the Kibble-Zurek argument, the scaling of the produced defect density depends on these critical exponents. 

The defects we are interested in here are excitations within the in-gap bands of the QWZ model. In this Appendix we calculate the corresponding critical exponents. We describe the evolution of the in-gap bands during a quench with the two-level Hamiltonian described with Eq.~\eqref{eq:H2}. The energy levels as a function of control parameter $u$ are shown in Fig.~\ref{fig:BandExc}(b). 

We extract the critical exponent $z\nu$ from the characteristic time scale, which is the inverse of the band gap. The spectrum near the gap closing is of the form $\pm\sqrt{\varepsilon^2+k_y^2}$, where $\varepsilon=u-u_c$ with $u_c=-1-\cos\frac{\pi}{N_x+1}$. The gap vanishes as $\varepsilon^{z\nu}=\varepsilon$, yielding the critical exponent $z\nu=1$.

For the calculation of the correlation length critical exponent $\nu$ we follow Ref.~\onlinecite{WeiChen16}. The correlation length is obtained from the scaling of the curvature function $F(\mathbf{k},\varepsilon)$ from which the winding number in $d$-dimensions $\mathcal{C}=\intop \mathrm{d}^d\mathbf{k}F(\mathbf{k},\varepsilon)$ is calculated. As the in-gap bands are one-dimensional, the winding number is calculated from the curvature function which is the Berry connection.\cite{Zak89} When rewriting the Hamiltonian~\eqref{eq:H2} as $\hat H_2(k_y)=\mathbf{d}(k_y)\cdot\hat{\boldsymbol{\sigma}}$, where $\mathbf{d}(k_y)=(0,\sin{k_y},\varepsilon+\cos{k_y}-1)$, the Berry connection can be calculated as $F(k_y,\varepsilon)=\frac{1}{2\pi}\big(\tilde{\mathbf{d}}(k_y)\times\partial_{k_y}\tilde{\mathbf{d}}(k_y)\big)_x$, where $\tilde{\mathbf{d}}=\mathbf{d}/|\mathbf{d}|$.\cite{madzari}
The length scale is then extracted from the scaling function,
\begin{equation}
\xi(\varepsilon) = \left|\frac{1}{\varepsilon}\frac{\partial^2_{k_y}F(k_y,\varepsilon)|_{k_y=0}}{\partial_\varepsilon F(0,\varepsilon)}\right|^\frac{1}{2},\label{eq:critL}
\end{equation}
yielding the critical exponent $\nu=1$.

\section{Time-averaged Hall conductance}
\label{app:HallDer}
\subsection{Perturbative evaluation of the Hall conductance}
\label{app:HallDer1}
In the main text we obtained the Hall conductance by calculating the Hall current in the state obtained by time evolution with a Hamiltonian which explicitly contained the time-dependent electric field. For small electric potentials one can evaluate the Hall conductance also perturbatively. Let the electrons immediately after a quench occupy the states $\{|\Phi_n(k_y) \rangle=\sum_{m}c_{nm}(k_y)|k_y\rangle\otimes|u_m(k_y)\rangle,\,1\leq n \leq N_x\}$ where $|u_m(k_y)\rangle$ are the eigenstates of the post-quench Hamiltonian $\hat{H}(k_y)$ without the electric field. Following Ref.~\onlinecite{Caio2016}, the response due to the electric potential $\hat{V}(t)=-qE_x(t)(\hat{x}-\frac{N_x+1}{2})$ can be evaluated using the time-dependent perturbation theory. As the electric potential is turned on, coefficients $c_{nm}(k_y)$ become time-dependent, $c_{nm}(k_y,t)=c_{nm}(k_y)+\delta c_{nm}(k_y,t)$. In the first order of perturbation the time-dependent part is equal to 
\begin{equation}
\begin{split}
\delta c_{nm}(k_y,t)=-i\sum_{p=1}^{2N_x}\intop_0^t\mathrm{d}t'\langle u_m(k_y)|\hat{V}(t')|u_p(k_y)\rangle\\ \times e^{i\Delta_{mp}(k_y)t'}c_{np}(k_y),
\label{deltac}
\end{split}
\end{equation}
where $\Delta_{mp}(k_y)=\varepsilon_m(k_y)-\varepsilon_p(k_y)$. We calculate the Hall current flowing in $y$-direction from Eq.~\eqref{eq:CurrOp}. To the first order in $E_x(t)$ it reads
\begin{equation}
\begin{split}
I_y(t)=2\mathrm{Re}\frac{1}{N_y}\sum_{k_y}\sum_{n=1}^{N_x}\sum_{m,p=1}^{2N_x}c^\ast_{nm}(k_y)\delta c_{np}(k_y,t)\times\\
e^{i\Delta_{mp}(k_y)t}\langle u_{m}(k_y)|q[\partial_{k_y}\hat{H}(k_y)]|u_p(k_y)\rangle.
\end{split}
\end{equation}
Inserting Eq.~\eqref{deltac} results in 
\begin{equation}
\begin{split}
&I_y(t)=2q^2E_0\,\mathrm{Re}\, \frac{1}{N_y}\sum_{ky}\sum_{n=1}^{N_x}\sum_{m,p,r=1}^{2N_x}\\
&c^{*}_{nm}(k_y)c_{nr}(k_y)f_{mpr}(k_y,t)\times\\
&\langle u_m(k_y)|[\partial_{k_y}\hat{H}(k_y)]|u_p(k_y)\rangle\times\\
&\langle u_p(k_y)|\hat{x}-\tfrac{N_x+1}{2}|u_r(k_y)\rangle\label{eq:IyTime}
\end{split}
\end{equation}
where we put all the time dependence into the function
\begin{equation}
f_{mpr}(k_y,t)=ie^{i\Delta_{mp}(k_y)t}\int_0^t\mathrm{d}t^\prime e^{i\Delta_{pr}(k_y)t^\prime}(1-e^{-t^\prime/\tau_E}).
\end{equation}
As we are interested in the long-time average of the current, we replace this function with $\bar f_{mpr}(k_y)=\lim_{T\rightarrow\infty}\frac{1}{T}\intop_0^T\mathrm{d}t f_{mpr}(k_y,t)$. The measurement of the Hall current unavoidably introduces decoherence and collapses the quenched state to a state represented by a diagonal ensemble,\cite{Rigol2008,LevRigol2016} i.e. the $m\ne r$ terms in Eq.~\eqref{eq:IyTime} die out. Of the remaining terms only those with $p\ne m$ contribute, $\bar f_{mpm}(k_y)=1/\Delta_{pm}(k_y)$. The $p=m$ term vanishes when the real part is taken in Eq.~\eqref{eq:IyTime} as $\bar f_{mmm}(k_y)$ is purely imaginary. The long-time average of the current is thus
\begin{equation}
\bar I_y=q^2E_0\frac{1}{N_y}\sum_{k_y}\sum_{m=1}^{2N_x}n_m(k_y)\Omega_m(k_y)
\end{equation}
where $n_m(k_y)=\sum_{n=1}^{N_x}|c_{nm}(k_y)|^2$ are post-quench occupancies of subbands and
\begin{equation}
\begin{split}
\Omega_{m}(k_y)=2\mathrm{Re}\sum_{p\neq m}\Delta_{pm}(k_y)^{-1}\times\\
\langle u_m(k_y)|[\partial_{k_y}\hat{H}(k_y)]|u_p(k_y)\rangle	\langle u_p(k_y)|\hat{x}|u_m(k_y)\rangle.
\end{split}
\end{equation}
Using the relation
$\langle u_m(k_y)|[\partial_{k_y}\hat{H}(k_y)]|u_p(k_y)\rangle=-\Delta_{pm}(k_y)\langle\partial_{k_y}u_m(k_y)|u_p(k_y)\rangle$ this simplifies to
\begin{equation}
\Omega_{m}(k_y)=-2\mathrm{Re}\sum_{p\neq m}\langle \partial_{k_y}u_m(k_y)|u_p(k_y)\rangle
\langle u_p(k_y)|\hat{x}|u_m(k_y)\rangle.
\label{eq:BerCurv1}
\end{equation}
Adding the purely imaginary $p=m$ term to the sum we end up with the expression from Eq.~\eqref{eq:BerryCurv},
\begin{equation}
\begin{split}
&\Omega_{m}(k_y)=-2\mathrm{Re}\langle\partial_{k_y}u_m(k_y)| \hat{x}|u_m(k_y)\rangle=\\
&-\partial_{k_y}\langle u_m(k_y)|\hat{x}|u_m(k_y)\rangle.\label{eq:BCurvApp}
\end{split}
\end{equation}

\begin{figure}[h]
	\centering
	\includegraphics[width=125pt]{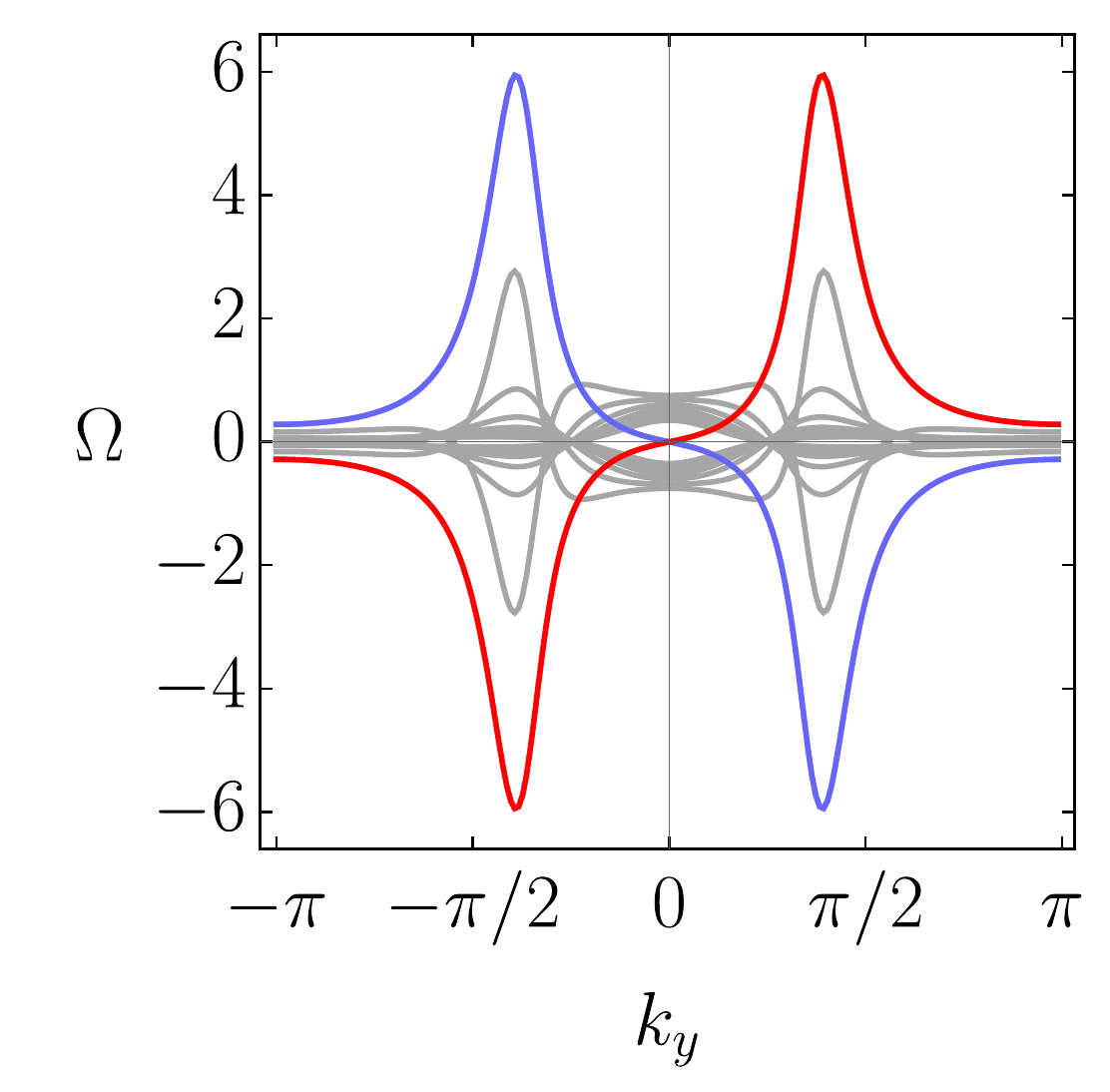}
	\caption{$\Omega_{L}(k_y)$ (blue) and $\Omega_{R}(k_y)$ (red) for a ribbon with $N_x=10$ at $u=-1.2$. $\Omega_m(k_y)$ for bulk subbands are plotted in gray color.}
	\label{fig:BCurv}
\end{figure}

$\Omega_m(k_y)$ at $u=-1.2$ are shown in Fig.~\ref{fig:BCurv}. Deep in the topological phase, i.e. for small $|u+1|$, $\Omega_m(k_y)$ for in-gap bands is linear in $k_y$ in the vicinity of $k_y=0$: $\Omega_{L,R}(k_y)=\pm2(u+1)k_y$.

\bibliography{bibliography_list}{}

\begin{thebibliography}{44}%
\makeatletter
\providecommand \@ifxundefined [1]{%
 \@ifx{#1\undefined}
}%
\providecommand \@ifnum [1]{%
 \ifnum #1\expandafter \@firstoftwo
 \else \expandafter \@secondoftwo
 \fi
}%
\providecommand \@ifx [1]{%
 \ifx #1\expandafter \@firstoftwo
 \else \expandafter \@secondoftwo
 \fi
}%
\providecommand \natexlab [1]{#1}%
\providecommand \enquote  [1]{``#1''}%
\providecommand \bibnamefont  [1]{#1}%
\providecommand \bibfnamefont [1]{#1}%
\providecommand \citenamefont [1]{#1}%
\providecommand \href@noop [0]{\@secondoftwo}%
\providecommand \href [0]{\begingroup \@sanitize@url \@href}%
\providecommand \@href[1]{\@@startlink{#1}\@@href}%
\providecommand \@@href[1]{\endgroup#1\@@endlink}%
\providecommand \@sanitize@url [0]{\catcode `\\12\catcode `\$12\catcode
  `\&12\catcode `\#12\catcode `\^12\catcode `\_12\catcode `\%12\relax}%
\providecommand \@@startlink[1]{}%
\providecommand \@@endlink[0]{}%
\providecommand \url  [0]{\begingroup\@sanitize@url \@url }%
\providecommand \@url [1]{\endgroup\@href {#1}{\urlprefix }}%
\providecommand \urlprefix  [0]{URL }%
\providecommand \Eprint [0]{\href }%
\providecommand \doibase [0]{http://dx.doi.org/}%
\providecommand \selectlanguage [0]{\@gobble}%
\providecommand \bibinfo  [0]{\@secondoftwo}%
\providecommand \bibfield  [0]{\@secondoftwo}%
\providecommand \translation [1]{[#1]}%
\providecommand \BibitemOpen [0]{}%
\providecommand \bibitemStop [0]{}%
\providecommand \bibitemNoStop [0]{.\EOS\space}%
\providecommand \EOS [0]{\spacefactor3000\relax}%
\providecommand \BibitemShut  [1]{\csname bibitem#1\endcsname}%
\let\auto@bib@innerbib\@empty
%</preamble>
\bibitem [{\citenamefont {Hasan}\ and\ \citenamefont {Kane}(2010)}]{Hasan10}%
  \BibitemOpen
  \bibfield  {author} {\bibinfo {author} {\bibfnamefont {M.~Z.}\ \bibnamefont
  {Hasan}}\ and\ \bibinfo {author} {\bibfnamefont {C.~L.}\ \bibnamefont
  {Kane}},\ }\href {\doibase 10.1103/RevModPhys.82.3045} {\bibfield  {journal}
  {\bibinfo  {journal} {Rev. Mod. Phys.}\ }\textbf {\bibinfo {volume} {82}},\
  \bibinfo {pages} {3045} (\bibinfo {year} {2010})}\BibitemShut {NoStop}%
\bibitem [{\citenamefont {Qi}\ and\ \citenamefont {Zhang}(2011)}]{Qi11}%
  \BibitemOpen
  \bibfield  {author} {\bibinfo {author} {\bibfnamefont {X.-L.}\ \bibnamefont
  {Qi}}\ and\ \bibinfo {author} {\bibfnamefont {S.-C.}\ \bibnamefont {Zhang}},\
  }\href {\doibase 10.1103/RevModPhys.83.1057} {\bibfield  {journal} {\bibinfo
  {journal} {Rev. Mod. Phys.}\ }\textbf {\bibinfo {volume} {83}},\ \bibinfo
  {pages} {1057} (\bibinfo {year} {2011})}\BibitemShut {NoStop}%
\bibitem [{\citenamefont {Tarruell}\ \emph {et~al.}(2012)\citenamefont
  {Tarruell}, \citenamefont {Greif}, \citenamefont {Uehlinger}, \citenamefont
  {Jotzu},\ and\ \citenamefont {Esslinger}}]{Tarruell12}%
  \BibitemOpen
  \bibfield  {author} {\bibinfo {author} {\bibfnamefont {L.}~\bibnamefont
  {Tarruell}}, \bibinfo {author} {\bibfnamefont {D.}~\bibnamefont {Greif}},
  \bibinfo {author} {\bibfnamefont {T.}~\bibnamefont {Uehlinger}}, \bibinfo
  {author} {\bibfnamefont {G.}~\bibnamefont {Jotzu}}, \ and\ \bibinfo {author}
  {\bibfnamefont {T.}~\bibnamefont {Esslinger}},\ }\href@noop {} {\bibfield
  {journal} {\bibinfo  {journal} {Nature}\ }\textbf {\bibinfo {volume} {483}},\
  \bibinfo {pages} {302} (\bibinfo {year} {2012})}\BibitemShut {NoStop}%
\bibitem [{\citenamefont {Aidelsburger}\ \emph {et~al.}(2013)\citenamefont
  {Aidelsburger}, \citenamefont {Atala}, \citenamefont {Lohse}, \citenamefont
  {Barreiro}, \citenamefont {Paredes},\ and\ \citenamefont
  {Bloch}}]{Aidelsburger13}%
  \BibitemOpen
  \bibfield  {author} {\bibinfo {author} {\bibfnamefont {M.}~\bibnamefont
  {Aidelsburger}}, \bibinfo {author} {\bibfnamefont {M.}~\bibnamefont {Atala}},
  \bibinfo {author} {\bibfnamefont {M.}~\bibnamefont {Lohse}}, \bibinfo
  {author} {\bibfnamefont {J.~T.}\ \bibnamefont {Barreiro}}, \bibinfo {author}
  {\bibfnamefont {B.}~\bibnamefont {Paredes}}, \ and\ \bibinfo {author}
  {\bibfnamefont {I.}~\bibnamefont {Bloch}},\ }\href {\doibase
  10.1103/PhysRevLett.111.185301} {\bibfield  {journal} {\bibinfo  {journal}
  {Phys. Rev. Lett.}\ }\textbf {\bibinfo {volume} {111}},\ \bibinfo {pages}
  {185301} (\bibinfo {year} {2013})}\BibitemShut {NoStop}%
\bibitem [{\citenamefont {Miyake}\ \emph {et~al.}(2013)\citenamefont {Miyake},
  \citenamefont {Siviloglou}, \citenamefont {Kennedy}, \citenamefont {Burton},\
  and\ \citenamefont {Ketterle}}]{Miyake13}%
  \BibitemOpen
  \bibfield  {author} {\bibinfo {author} {\bibfnamefont {H.}~\bibnamefont
  {Miyake}}, \bibinfo {author} {\bibfnamefont {G.~A.}\ \bibnamefont
  {Siviloglou}}, \bibinfo {author} {\bibfnamefont {C.~J.}\ \bibnamefont
  {Kennedy}}, \bibinfo {author} {\bibfnamefont {W.~C.}\ \bibnamefont {Burton}},
  \ and\ \bibinfo {author} {\bibfnamefont {W.}~\bibnamefont {Ketterle}},\
  }\href {\doibase 10.1103/PhysRevLett.111.185302} {\bibfield  {journal}
  {\bibinfo  {journal} {Phys. Rev. Lett.}\ }\textbf {\bibinfo {volume} {111}},\
  \bibinfo {pages} {185302} (\bibinfo {year} {2013})}\BibitemShut {NoStop}%
\bibitem [{\citenamefont {Goldman}\ \emph {et~al.}(2013)\citenamefont
  {Goldman}, \citenamefont {Anisimovas}, \citenamefont {Gerbier}, \citenamefont
  {Öhberg}, \citenamefont {Spielman},\ and\ \citenamefont
  {Juzeli{\={u}}nas}}]{Goldman13}%
  \BibitemOpen
  \bibfield  {author} {\bibinfo {author} {\bibfnamefont {N.}~\bibnamefont
  {Goldman}}, \bibinfo {author} {\bibfnamefont {E.}~\bibnamefont {Anisimovas}},
  \bibinfo {author} {\bibfnamefont {F.}~\bibnamefont {Gerbier}}, \bibinfo
  {author} {\bibfnamefont {P.}~\bibnamefont {Öhberg}}, \bibinfo {author}
  {\bibfnamefont {I.~B.}\ \bibnamefont {Spielman}}, \ and\ \bibinfo {author}
  {\bibfnamefont {G.}~\bibnamefont {Juzeli{\={u}}nas}},\ }\href {\doibase
  10.1088/1367-2630/15/1/013025} {\bibfield  {journal} {\bibinfo  {journal}
  {New Journal of Physics}\ }\textbf {\bibinfo {volume} {15}},\ \bibinfo
  {pages} {013025} (\bibinfo {year} {2013})}\BibitemShut {NoStop}%
\bibitem [{\citenamefont {Dauphin}\ and\ \citenamefont
  {Goldman}(2013)}]{Dauphin13}%
  \BibitemOpen
  \bibfield  {author} {\bibinfo {author} {\bibfnamefont {A.}~\bibnamefont
  {Dauphin}}\ and\ \bibinfo {author} {\bibfnamefont {N.}~\bibnamefont
  {Goldman}},\ }\href {\doibase 10.1103/PhysRevLett.111.135302} {\bibfield
  {journal} {\bibinfo  {journal} {Phys. Rev. Lett.}\ }\textbf {\bibinfo
  {volume} {111}},\ \bibinfo {pages} {135302} (\bibinfo {year}
  {2013})}\BibitemShut {NoStop}%
\bibitem [{\citenamefont {Jotzu}\ \emph {et~al.}(2014)\citenamefont {Jotzu},
  \citenamefont {Messer}, \citenamefont {Desbuquois}, \citenamefont {Lebrat},
  \citenamefont {Uehlinger}, \citenamefont {Greif},\ and\ \citenamefont
  {Esslinger}}]{Jotzu14}%
  \BibitemOpen
  \bibfield  {author} {\bibinfo {author} {\bibfnamefont {G.}~\bibnamefont
  {Jotzu}}, \bibinfo {author} {\bibfnamefont {M.}~\bibnamefont {Messer}},
  \bibinfo {author} {\bibfnamefont {R.}~\bibnamefont {Desbuquois}}, \bibinfo
  {author} {\bibfnamefont {M.}~\bibnamefont {Lebrat}}, \bibinfo {author}
  {\bibfnamefont {T.}~\bibnamefont {Uehlinger}}, \bibinfo {author}
  {\bibfnamefont {D.}~\bibnamefont {Greif}}, \ and\ \bibinfo {author}
  {\bibfnamefont {T.}~\bibnamefont {Esslinger}},\ }\href {\doibase
  10.1038/nature13915} {\bibfield  {journal} {\bibinfo  {journal} {Nature}\
  }\textbf {\bibinfo {volume} {515}},\ \bibinfo {pages} {237} (\bibinfo {year}
  {2014})}\BibitemShut {NoStop}%
\bibitem [{\citenamefont {Wu}\ \emph {et~al.}(2016)\citenamefont {Wu},
  \citenamefont {Zhang}, \citenamefont {Sun}, \citenamefont {Xu}, \citenamefont
  {Wang}, \citenamefont {Ji}, \citenamefont {Deng}, \citenamefont {Chen},
  \citenamefont {Liu},\ and\ \citenamefont {Pan}}]{Wu16}%
  \BibitemOpen
  \bibfield  {author} {\bibinfo {author} {\bibfnamefont {Z.}~\bibnamefont
  {Wu}}, \bibinfo {author} {\bibfnamefont {L.}~\bibnamefont {Zhang}}, \bibinfo
  {author} {\bibfnamefont {W.}~\bibnamefont {Sun}}, \bibinfo {author}
  {\bibfnamefont {X.-T.}\ \bibnamefont {Xu}}, \bibinfo {author} {\bibfnamefont
  {B.-Z.}\ \bibnamefont {Wang}}, \bibinfo {author} {\bibfnamefont {S.-C.}\
  \bibnamefont {Ji}}, \bibinfo {author} {\bibfnamefont {Y.}~\bibnamefont
  {Deng}}, \bibinfo {author} {\bibfnamefont {S.}~\bibnamefont {Chen}}, \bibinfo
  {author} {\bibfnamefont {X.-J.}\ \bibnamefont {Liu}}, \ and\ \bibinfo
  {author} {\bibfnamefont {J.-W.}\ \bibnamefont {Pan}},\ }\href {\doibase
  10.1126/science.aaf6689} {\bibfield  {journal} {\bibinfo  {journal}
  {Science}\ }\textbf {\bibinfo {volume} {354}},\ \bibinfo {pages} {83}
  (\bibinfo {year} {2016})}\BibitemShut {NoStop}%
\bibitem [{\citenamefont {Fl{\"a}schner}\ \emph {et~al.}(2016)\citenamefont
  {Fl{\"a}schner}, \citenamefont {Rem}, \citenamefont {Tarnowski},
  \citenamefont {Vogel}, \citenamefont {L{\"u}hmann}, \citenamefont
  {Sengstock},\ and\ \citenamefont {Weitenberg}}]{Flaschner16}%
  \BibitemOpen
  \bibfield  {author} {\bibinfo {author} {\bibfnamefont {N.}~\bibnamefont
  {Fl{\"a}schner}}, \bibinfo {author} {\bibfnamefont {B.~S.}\ \bibnamefont
  {Rem}}, \bibinfo {author} {\bibfnamefont {M.}~\bibnamefont {Tarnowski}},
  \bibinfo {author} {\bibfnamefont {D.}~\bibnamefont {Vogel}}, \bibinfo
  {author} {\bibfnamefont {D.-S.}\ \bibnamefont {L{\"u}hmann}}, \bibinfo
  {author} {\bibfnamefont {K.}~\bibnamefont {Sengstock}}, \ and\ \bibinfo
  {author} {\bibfnamefont {C.}~\bibnamefont {Weitenberg}},\ }\href {\doibase
  10.1126/science.aad4568} {\bibfield  {journal} {\bibinfo  {journal}
  {Science}\ }\textbf {\bibinfo {volume} {352}},\ \bibinfo {pages} {1091}
  (\bibinfo {year} {2016})}\BibitemShut {NoStop}%
\bibitem [{\citenamefont {K\"{o}nig}\ \emph {et~al.}(2007)\citenamefont
  {K\"{o}nig}, \citenamefont {Br\"{u}ne}, \citenamefont {Roth}, \citenamefont
  {Buhmann}, \citenamefont {Molenkamp}, \citenamefont {Qi},\ and\ \citenamefont
  {Zhang}}]{HgTeExp}%
  \BibitemOpen
  \bibfield  {author} {\bibinfo {author} {\bibfnamefont {W.~S.}\ \bibnamefont
  {K\"{o}nig}, \bibfnamefont {Markus}}, \bibinfo {author} {\bibfnamefont
  {C.}~\bibnamefont {Br\"{u}ne}}, \bibinfo {author} {\bibfnamefont
  {A.}~\bibnamefont {Roth}}, \bibinfo {author} {\bibfnamefont {H.}~\bibnamefont
  {Buhmann}}, \bibinfo {author} {\bibfnamefont {L.~W.}\ \bibnamefont
  {Molenkamp}}, \bibinfo {author} {\bibfnamefont {X.-L.}\ \bibnamefont {Qi}}, \
  and\ \bibinfo {author} {\bibfnamefont {S.-C.}\ \bibnamefont {Zhang}},\ }\href
  {\doibase 10.1126/science.1148047} {\bibfield  {journal} {\bibinfo  {journal}
  {Science}\ }\textbf {\bibinfo {volume} {318}},\ \bibinfo {pages} {766}
  (\bibinfo {year} {2007})}\BibitemShut {NoStop}%
\bibitem [{\citenamefont {Yu}\ \emph {et~al.}(2010)\citenamefont {Yu},
  \citenamefont {Zhang}, \citenamefont {Zhang}, \citenamefont {Zhang},
  \citenamefont {Dai},\ and\ \citenamefont {Fang}}]{Yu10}%
  \BibitemOpen
  \bibfield  {author} {\bibinfo {author} {\bibfnamefont {R.}~\bibnamefont
  {Yu}}, \bibinfo {author} {\bibfnamefont {W.}~\bibnamefont {Zhang}}, \bibinfo
  {author} {\bibfnamefont {H.-J.}\ \bibnamefont {Zhang}}, \bibinfo {author}
  {\bibfnamefont {S.-C.}\ \bibnamefont {Zhang}}, \bibinfo {author}
  {\bibfnamefont {X.}~\bibnamefont {Dai}}, \ and\ \bibinfo {author}
  {\bibfnamefont {Z.}~\bibnamefont {Fang}},\ }\href {\doibase
  10.1126/science.1187485} {\bibfield  {journal} {\bibinfo  {journal}
  {Science}\ }\textbf {\bibinfo {volume} {329}},\ \bibinfo {pages} {61}
  (\bibinfo {year} {2010})}\BibitemShut {NoStop}%
\bibitem [{\citenamefont {Knez}\ \emph {et~al.}(2011)\citenamefont {Knez},
  \citenamefont {Du},\ and\ \citenamefont {Sullivan}}]{InAsExp}%
  \BibitemOpen
  \bibfield  {author} {\bibinfo {author} {\bibfnamefont {I.}~\bibnamefont
  {Knez}}, \bibinfo {author} {\bibfnamefont {R.-R.}\ \bibnamefont {Du}}, \ and\
  \bibinfo {author} {\bibfnamefont {G.}~\bibnamefont {Sullivan}},\ }\href
  {\doibase 10.1103/PhysRevLett.107.136603} {\bibfield  {journal} {\bibinfo
  {journal} {Phys. Rev. Lett.}\ }\textbf {\bibinfo {volume} {107}},\ \bibinfo
  {pages} {136603} (\bibinfo {year} {2011})}\BibitemShut {NoStop}%
\bibitem [{\citenamefont {Chang}\ \emph {et~al.}(2013)\citenamefont {Chang},
  \citenamefont {Zhang}, \citenamefont {Feng}, \citenamefont {Shen},
  \citenamefont {Zhang}, \citenamefont {Guo}, \citenamefont {Li}, \citenamefont
  {Ou}, \citenamefont {Wei}, \citenamefont {Wang}, \citenamefont {Ji},
  \citenamefont {Feng}, \citenamefont {Ji}, \citenamefont {Chen}, \citenamefont
  {Jia}, \citenamefont {Dai}, \citenamefont {Fang}, \citenamefont {Zhang},
  \citenamefont {He}, \citenamefont {Wang}, \citenamefont {Lu}, \citenamefont
  {Ma},\ and\ \citenamefont {Xue}}]{Chang13}%
  \BibitemOpen
  \bibfield  {author} {\bibinfo {author} {\bibfnamefont {C.-Z.}\ \bibnamefont
  {Chang}}, \bibinfo {author} {\bibfnamefont {J.}~\bibnamefont {Zhang}},
  \bibinfo {author} {\bibfnamefont {X.}~\bibnamefont {Feng}}, \bibinfo {author}
  {\bibfnamefont {J.}~\bibnamefont {Shen}}, \bibinfo {author} {\bibfnamefont
  {Z.}~\bibnamefont {Zhang}}, \bibinfo {author} {\bibfnamefont
  {M.}~\bibnamefont {Guo}}, \bibinfo {author} {\bibfnamefont {K.}~\bibnamefont
  {Li}}, \bibinfo {author} {\bibfnamefont {Y.}~\bibnamefont {Ou}}, \bibinfo
  {author} {\bibfnamefont {P.}~\bibnamefont {Wei}}, \bibinfo {author}
  {\bibfnamefont {L.-L.}\ \bibnamefont {Wang}}, \bibinfo {author}
  {\bibfnamefont {Z.-Q.}\ \bibnamefont {Ji}}, \bibinfo {author} {\bibfnamefont
  {Y.}~\bibnamefont {Feng}}, \bibinfo {author} {\bibfnamefont {S.}~\bibnamefont
  {Ji}}, \bibinfo {author} {\bibfnamefont {X.}~\bibnamefont {Chen}}, \bibinfo
  {author} {\bibfnamefont {J.}~\bibnamefont {Jia}}, \bibinfo {author}
  {\bibfnamefont {X.}~\bibnamefont {Dai}}, \bibinfo {author} {\bibfnamefont
  {Z.}~\bibnamefont {Fang}}, \bibinfo {author} {\bibfnamefont {S.-C.}\
  \bibnamefont {Zhang}}, \bibinfo {author} {\bibfnamefont {K.}~\bibnamefont
  {He}}, \bibinfo {author} {\bibfnamefont {Y.}~\bibnamefont {Wang}}, \bibinfo
  {author} {\bibfnamefont {L.}~\bibnamefont {Lu}}, \bibinfo {author}
  {\bibfnamefont {X.-C.}\ \bibnamefont {Ma}}, \ and\ \bibinfo {author}
  {\bibfnamefont {Q.-K.}\ \bibnamefont {Xue}},\ }\href {\doibase
  10.1126/science.1234414} {\bibfield  {journal} {\bibinfo  {journal}
  {Science}\ }\textbf {\bibinfo {volume} {340}},\ \bibinfo {pages} {167}
  (\bibinfo {year} {2013})}\BibitemShut {NoStop}%
\bibitem [{\citenamefont {Reis}\ \emph {et~al.}(2017)\citenamefont {Reis},
  \citenamefont {Li}, \citenamefont {Dudy}, \citenamefont {Bauernfeind},
  \citenamefont {Glass}, \citenamefont {Hanke}, \citenamefont {Thomale},
  \citenamefont {Sch{\"a}fer},\ and\ \citenamefont {Claessen}}]{Bismuthene}%
  \BibitemOpen
  \bibfield  {author} {\bibinfo {author} {\bibfnamefont {F.}~\bibnamefont
  {Reis}}, \bibinfo {author} {\bibfnamefont {G.}~\bibnamefont {Li}}, \bibinfo
  {author} {\bibfnamefont {L.}~\bibnamefont {Dudy}}, \bibinfo {author}
  {\bibfnamefont {M.}~\bibnamefont {Bauernfeind}}, \bibinfo {author}
  {\bibfnamefont {S.}~\bibnamefont {Glass}}, \bibinfo {author} {\bibfnamefont
  {W.}~\bibnamefont {Hanke}}, \bibinfo {author} {\bibfnamefont
  {R.}~\bibnamefont {Thomale}}, \bibinfo {author} {\bibfnamefont
  {J.}~\bibnamefont {Sch{\"a}fer}}, \ and\ \bibinfo {author} {\bibfnamefont
  {R.}~\bibnamefont {Claessen}},\ }\href {\doibase 10.1126/science.aai8142}
  {\bibfield  {journal} {\bibinfo  {journal} {Science}\ }\textbf {\bibinfo
  {volume} {357}},\ \bibinfo {pages} {287} (\bibinfo {year} {2017})},\ \Eprint
  {http://arxiv.org/abs/http://science.sciencemag.org/content/357/6348/287.full.pdf}
  {http://science.sciencemag.org/content/357/6348/287.full.pdf} \BibitemShut
  {NoStop}%
\bibitem [{\citenamefont {Tokura}\ \emph {et~al.}(2019)\citenamefont {Tokura},
  \citenamefont {Yasuda},\ and\ \citenamefont {Tsukazaki}}]{Tokura19}%
  \BibitemOpen
  \bibfield  {author} {\bibinfo {author} {\bibfnamefont {Y.}~\bibnamefont
  {Tokura}}, \bibinfo {author} {\bibfnamefont {K.}~\bibnamefont {Yasuda}}, \
  and\ \bibinfo {author} {\bibfnamefont {A.}~\bibnamefont {Tsukazaki}},\
  }\href@noop {} {\bibfield  {journal} {\bibinfo  {journal} {Nature Reviews
  Physics}\ }\textbf {\bibinfo {volume} {1}},\ \bibinfo {pages} {126} (\bibinfo
  {year} {2019})}\BibitemShut {NoStop}%
\bibitem [{\citenamefont {Caio}\ \emph {et~al.}(2016)\citenamefont {Caio},
  \citenamefont {Cooper},\ and\ \citenamefont {Bhaseen}}]{Caio2016}%
  \BibitemOpen
  \bibfield  {author} {\bibinfo {author} {\bibfnamefont {M.~D.}\ \bibnamefont
  {Caio}}, \bibinfo {author} {\bibfnamefont {N.~R.}\ \bibnamefont {Cooper}}, \
  and\ \bibinfo {author} {\bibfnamefont {M.~J.}\ \bibnamefont {Bhaseen}},\
  }\href {\doibase 10.1103/PhysRevB.94.155104} {\bibfield  {journal} {\bibinfo
  {journal} {Phys. Rev. B}\ }\textbf {\bibinfo {volume} {94}},\ \bibinfo
  {pages} {155104} (\bibinfo {year} {2016})}\BibitemShut {NoStop}%
\bibitem [{\citenamefont {\"Unal}\ \emph {et~al.}(2016)\citenamefont {\"Unal},
  \citenamefont {Mueller},\ and\ \citenamefont {Oktel}}]{Unal16}%
  \BibitemOpen
  \bibfield  {author} {\bibinfo {author} {\bibfnamefont {F.~N.}\ \bibnamefont
  {\"Unal}}, \bibinfo {author} {\bibfnamefont {E.~J.}\ \bibnamefont {Mueller}},
  \ and\ \bibinfo {author} {\bibfnamefont {M.~O.}\ \bibnamefont {Oktel}},\
  }\href {\doibase 10.1103/PhysRevA.94.053604} {\bibfield  {journal} {\bibinfo
  {journal} {Phys. Rev. A}\ }\textbf {\bibinfo {volume} {94}},\ \bibinfo
  {pages} {053604} (\bibinfo {year} {2016})}\BibitemShut {NoStop}%
\bibitem [{\citenamefont {Hu}\ \emph {et~al.}(2016)\citenamefont {Hu},
  \citenamefont {Zoller},\ and\ \citenamefont {Budich}}]{zoller}%
  \BibitemOpen
  \bibfield  {author} {\bibinfo {author} {\bibfnamefont {Y.}~\bibnamefont
  {Hu}}, \bibinfo {author} {\bibfnamefont {P.}~\bibnamefont {Zoller}}, \ and\
  \bibinfo {author} {\bibfnamefont {J.~C.}\ \bibnamefont {Budich}},\ }\href
  {\doibase 10.1103/PhysRevLett.117.126803} {\bibfield  {journal} {\bibinfo
  {journal} {Phys. Rev. Lett.}\ }\textbf {\bibinfo {volume} {117}},\ \bibinfo
  {pages} {126803} (\bibinfo {year} {2016})}\BibitemShut {NoStop}%
\bibitem [{\citenamefont {Ul\v{c}akar}\ \emph {et~al.}(2018)\citenamefont
  {Ul\v{c}akar}, \citenamefont {Mravlje}, \citenamefont {Ram\v{s}ak},\ and\
  \citenamefont {Rejec}}]{Ulcakar2018}%
  \BibitemOpen
  \bibfield  {author} {\bibinfo {author} {\bibfnamefont {L.}~\bibnamefont
  {Ul\v{c}akar}}, \bibinfo {author} {\bibfnamefont {J.}~\bibnamefont
  {Mravlje}}, \bibinfo {author} {\bibfnamefont {A.}~\bibnamefont {Ram\v{s}ak}},
  \ and\ \bibinfo {author} {\bibfnamefont {T.}~\bibnamefont {Rejec}},\ }\href
  {\doibase 10.1103/PhysRevB.97.195127} {\bibfield  {journal} {\bibinfo
  {journal} {Phys. Rev. B}\ }\textbf {\bibinfo {volume} {97}},\ \bibinfo
  {pages} {195127} (\bibinfo {year} {2018})}\BibitemShut {NoStop}%
\bibitem [{\citenamefont {Damski}(2005)}]{Damski05}%
  \BibitemOpen
  \bibfield  {author} {\bibinfo {author} {\bibfnamefont {B.}~\bibnamefont
  {Damski}},\ }\href {\doibase 10.1103/PhysRevLett.95.035701} {\bibfield
  {journal} {\bibinfo  {journal} {Phys. Rev. Lett.}\ }\textbf {\bibinfo
  {volume} {95}},\ \bibinfo {pages} {035701} (\bibinfo {year}
  {2005})}\BibitemShut {NoStop}%
\bibitem [{\citenamefont {Dutta}\ \emph {et~al.}(2010)\citenamefont {Dutta},
  \citenamefont {Singh},\ and\ \citenamefont {Divakaran}}]{Dutta10}%
  \BibitemOpen
  \bibfield  {author} {\bibinfo {author} {\bibfnamefont {A.}~\bibnamefont
  {Dutta}}, \bibinfo {author} {\bibfnamefont {R.~R.~P.}\ \bibnamefont {Singh}},
  \ and\ \bibinfo {author} {\bibfnamefont {U.}~\bibnamefont {Divakaran}},\
  }\href {http://stacks.iop.org/0295-5075/89/i=6/a=67001} {\bibfield  {journal}
  {\bibinfo  {journal} {EPL (Europhysics Letters)}\ }\textbf {\bibinfo {volume}
  {89}},\ \bibinfo {pages} {67001} (\bibinfo {year} {2010})}\BibitemShut
  {NoStop}%
\bibitem [{\citenamefont {Kibble}(1976)}]{Kibble}%
  \BibitemOpen
  \bibfield  {author} {\bibinfo {author} {\bibfnamefont {T.~W.~B.}\
  \bibnamefont {Kibble}},\ }\href {http://stacks.iop.org/0305-4470/9/i=8/a=029}
  {\bibfield  {journal} {\bibinfo  {journal} {Journal of Physics A:
  Mathematical and General}\ }\textbf {\bibinfo {volume} {9}},\ \bibinfo
  {pages} {1387} (\bibinfo {year} {1976})}\BibitemShut {NoStop}%
\bibitem [{\citenamefont {Zurek}(1985)}]{Zurek}%
  \BibitemOpen
  \bibfield  {author} {\bibinfo {author} {\bibfnamefont {W.~H.}\ \bibnamefont
  {Zurek}},\ }\href@noop {} {\bibfield  {journal} {\bibinfo  {journal}
  {Nature}\ }\textbf {\bibinfo {volume} {317}},\ \bibinfo {pages} {505}
  (\bibinfo {year} {1985})}\BibitemShut {NoStop}%
\bibitem [{\citenamefont {McGinley}\ and\ \citenamefont
  {Cooper}(2019)}]{McGinley18}%
  \BibitemOpen
  \bibfield  {author} {\bibinfo {author} {\bibfnamefont {M.}~\bibnamefont
  {McGinley}}\ and\ \bibinfo {author} {\bibfnamefont {N.~R.}\ \bibnamefont
  {Cooper}},\ }\href {\doibase 10.1103/PhysRevB.99.075148} {\bibfield
  {journal} {\bibinfo  {journal} {Phys. Rev. B}\ }\textbf {\bibinfo {volume}
  {99}},\ \bibinfo {pages} {075148} (\bibinfo {year} {2019})}\BibitemShut
  {NoStop}%
\bibitem [{\citenamefont {D'’Alessio}\ and\ \citenamefont
  {Rigol}(2015)}]{Rigol2015}%
  \BibitemOpen
  \bibfield  {author} {\bibinfo {author} {\bibfnamefont {L.}~\bibnamefont
  {D'’Alessio}}\ and\ \bibinfo {author} {\bibfnamefont {M.}~\bibnamefont
  {Rigol}},\ }\href@noop {} {\bibfield  {journal} {\bibinfo  {journal} {Nature
  Communications}\ }\textbf {\bibinfo {volume} {6}},\ \bibinfo {pages} {8336}
  (\bibinfo {year} {2015})}\BibitemShut {NoStop}%
\bibitem [{\citenamefont {Caio}\ \emph {et~al.}(2015)\citenamefont {Caio},
  \citenamefont {Cooper},\ and\ \citenamefont {Bhaseen}}]{Caio2015}%
  \BibitemOpen
  \bibfield  {author} {\bibinfo {author} {\bibfnamefont {M.~D.}\ \bibnamefont
  {Caio}}, \bibinfo {author} {\bibfnamefont {N.~R.}\ \bibnamefont {Cooper}}, \
  and\ \bibinfo {author} {\bibfnamefont {M.~J.}\ \bibnamefont {Bhaseen}},\
  }\href {\doibase 10.1103/PhysRevLett.115.236403} {\bibfield  {journal}
  {\bibinfo  {journal} {Phys. Rev. Lett.}\ }\textbf {\bibinfo {volume} {115}},\
  \bibinfo {pages} {236403} (\bibinfo {year} {2015})}\BibitemShut {NoStop}%
\bibitem [{\citenamefont {Foster}\ \emph {et~al.}(2013)\citenamefont {Foster},
  \citenamefont {Dzero}, \citenamefont {Gurarie},\ and\ \citenamefont
  {Yuzbashyan}}]{PxPySuperFluid}%
  \BibitemOpen
  \bibfield  {author} {\bibinfo {author} {\bibfnamefont {M.~S.}\ \bibnamefont
  {Foster}}, \bibinfo {author} {\bibfnamefont {M.}~\bibnamefont {Dzero}},
  \bibinfo {author} {\bibfnamefont {V.}~\bibnamefont {Gurarie}}, \ and\
  \bibinfo {author} {\bibfnamefont {E.~A.}\ \bibnamefont {Yuzbashyan}},\ }\href
  {\doibase 10.1103/PhysRevB.88.104511} {\bibfield  {journal} {\bibinfo
  {journal} {Phys. Rev. B}\ }\textbf {\bibinfo {volume} {88}},\ \bibinfo
  {pages} {104511} (\bibinfo {year} {2013})}\BibitemShut {NoStop}%
\bibitem [{\citenamefont {Bhattacharya}\ \emph {et~al.}(2017)\citenamefont
  {Bhattacharya}, \citenamefont {Hutchinson},\ and\ \citenamefont
  {Dutta}}]{Duta17}%
  \BibitemOpen
  \bibfield  {author} {\bibinfo {author} {\bibfnamefont {U.}~\bibnamefont
  {Bhattacharya}}, \bibinfo {author} {\bibfnamefont {J.}~\bibnamefont
  {Hutchinson}}, \ and\ \bibinfo {author} {\bibfnamefont {A.}~\bibnamefont
  {Dutta}},\ }\href {\doibase 10.1103/PhysRevB.95.144304} {\bibfield  {journal}
  {\bibinfo  {journal} {Phys. Rev. B}\ }\textbf {\bibinfo {volume} {95}},\
  \bibinfo {pages} {144304} (\bibinfo {year} {2017})}\BibitemShut {NoStop}%
\bibitem [{\citenamefont {Dehghani}\ and\ \citenamefont
  {Mitra}(2016)}]{Mitra16}%
  \BibitemOpen
  \bibfield  {author} {\bibinfo {author} {\bibfnamefont {H.}~\bibnamefont
  {Dehghani}}\ and\ \bibinfo {author} {\bibfnamefont {A.}~\bibnamefont
  {Mitra}},\ }\href {\doibase 10.1103/PhysRevB.93.205437} {\bibfield  {journal}
  {\bibinfo  {journal} {Phys. Rev. B}\ }\textbf {\bibinfo {volume} {93}},\
  \bibinfo {pages} {205437} (\bibinfo {year} {2016})}\BibitemShut {NoStop}%
\bibitem [{\citenamefont {Privitera}\ and\ \citenamefont
  {Santoro}(2016)}]{Privitera16}%
  \BibitemOpen
  \bibfield  {author} {\bibinfo {author} {\bibfnamefont {L.}~\bibnamefont
  {Privitera}}\ and\ \bibinfo {author} {\bibfnamefont {G.~E.}\ \bibnamefont
  {Santoro}},\ }\href {\doibase 10.1103/PhysRevB.93.241406} {\bibfield
  {journal} {\bibinfo  {journal} {Phys. Rev. B}\ }\textbf {\bibinfo {volume}
  {93}},\ \bibinfo {pages} {241406} (\bibinfo {year} {2016})}\BibitemShut
  {NoStop}%
\bibitem [{\citenamefont {Haldane}(1988)}]{HaldaneModel}%
  \BibitemOpen
  \bibfield  {author} {\bibinfo {author} {\bibfnamefont {F.~D.~M.}\
  \bibnamefont {Haldane}},\ }\href {\doibase 10.1103/PhysRevLett.61.2015}
  {\bibfield  {journal} {\bibinfo  {journal} {Phys. Rev. Lett.}\ }\textbf
  {\bibinfo {volume} {61}},\ \bibinfo {pages} {2015} (\bibinfo {year}
  {1988})}\BibitemShut {NoStop}%
\bibitem [{\citenamefont {Bermudez}\ \emph {et~al.}(2009)\citenamefont
  {Bermudez}, \citenamefont {Patan\`e}, \citenamefont {Amico},\ and\
  \citenamefont {Martin-Delgado}}]{Bermudez09}%
  \BibitemOpen
  \bibfield  {author} {\bibinfo {author} {\bibfnamefont {A.}~\bibnamefont
  {Bermudez}}, \bibinfo {author} {\bibfnamefont {D.}~\bibnamefont {Patan\`e}},
  \bibinfo {author} {\bibfnamefont {L.}~\bibnamefont {Amico}}, \ and\ \bibinfo
  {author} {\bibfnamefont {M.~A.}\ \bibnamefont {Martin-Delgado}},\ }\href
  {\doibase 10.1103/PhysRevLett.102.135702} {\bibfield  {journal} {\bibinfo
  {journal} {Phys. Rev. Lett.}\ }\textbf {\bibinfo {volume} {102}},\ \bibinfo
  {pages} {135702} (\bibinfo {year} {2009})}\BibitemShut {NoStop}%
\bibitem [{\citenamefont {Bermudez}\ \emph {et~al.}(2010)\citenamefont
  {Bermudez}, \citenamefont {Amico},\ and\ \citenamefont
  {Martin-Delgado}}]{Bermudez10}%
  \BibitemOpen
  \bibfield  {author} {\bibinfo {author} {\bibfnamefont {A.}~\bibnamefont
  {Bermudez}}, \bibinfo {author} {\bibfnamefont {L.}~\bibnamefont {Amico}}, \
  and\ \bibinfo {author} {\bibfnamefont {M.~A.}\ \bibnamefont
  {Martin-Delgado}},\ }\href {http://stacks.iop.org/1367-2630/12/i=5/a=055014}
  {\bibfield  {journal} {\bibinfo  {journal} {New Journal of Physics}\ }\textbf
  {\bibinfo {volume} {12}},\ \bibinfo {pages} {055014} (\bibinfo {year}
  {2010})}\BibitemShut {NoStop}%
\bibitem [{\citenamefont {Qi}\ \emph {et~al.}(2006)\citenamefont {Qi},
  \citenamefont {Wu},\ and\ \citenamefont {Zhang}}]{QWZmodel}%
  \BibitemOpen
  \bibfield  {author} {\bibinfo {author} {\bibfnamefont {X.-L.}\ \bibnamefont
  {Qi}}, \bibinfo {author} {\bibfnamefont {Y.-S.}\ \bibnamefont {Wu}}, \ and\
  \bibinfo {author} {\bibfnamefont {S.-C.}\ \bibnamefont {Zhang}},\ }\href
  {\doibase 10.1103/PhysRevB.74.085308} {\bibfield  {journal} {\bibinfo
  {journal} {Phys. Rev. B}\ }\textbf {\bibinfo {volume} {74}},\ \bibinfo
  {pages} {085308} (\bibinfo {year} {2006})}\BibitemShut {NoStop}%
\bibitem [{\citenamefont {Bernevig}\ \emph {et~al.}(2006)\citenamefont
  {Bernevig}, \citenamefont {Hughes},\ and\ \citenamefont {Zhang}}]{BHZmodel}%
  \BibitemOpen
  \bibfield  {author} {\bibinfo {author} {\bibfnamefont {B.~A.}\ \bibnamefont
  {Bernevig}}, \bibinfo {author} {\bibfnamefont {T.~L.}\ \bibnamefont
  {Hughes}}, \ and\ \bibinfo {author} {\bibfnamefont {S.-C.}\ \bibnamefont
  {Zhang}},\ }\href {\doibase 10.1126/science.1133734} {\bibfield  {journal}
  {\bibinfo  {journal} {Science}\ }\textbf {\bibinfo {volume} {314}},\ \bibinfo
  {pages} {1757} (\bibinfo {year} {2006})}\BibitemShut {NoStop}%
\bibitem [{\citenamefont {Chen}(2016)}]{Chen16}%
  \BibitemOpen
  \bibfield  {author} {\bibinfo {author} {\bibfnamefont {W.}~\bibnamefont
  {Chen}},\ }\href {\doibase 10.1088/0953-8984/28/5/055601} {\bibfield
  {journal} {\bibinfo  {journal} {Journal of Physics: Condensed Matter}\
  }\textbf {\bibinfo {volume} {28}},\ \bibinfo {pages} {055601} (\bibinfo
  {year} {2016})}\BibitemShut {NoStop}%
\bibitem [{\citenamefont {Asb\'{o}th}\ \emph {et~al.}(2016)\citenamefont
  {Asb\'{o}th}, \citenamefont {Oroszl\'{a}ny},\ and\ \citenamefont
  {P\'{a}lyi}}]{madzari}%
  \BibitemOpen
  \bibfield  {author} {\bibinfo {author} {\bibfnamefont {J.~K.}\ \bibnamefont
  {Asb\'{o}th}}, \bibinfo {author} {\bibfnamefont {L.}~\bibnamefont
  {Oroszl\'{a}ny}}, \ and\ \bibinfo {author} {\bibfnamefont {A.}~\bibnamefont
  {P\'{a}lyi}},\ }\href@noop {} {\emph {\bibinfo {title} {A Short Course on
  Topological Insulators}}}\ (\bibinfo  {publisher} {Springer},\ \bibinfo
  {year} {2016})\BibitemShut {NoStop}%
\bibitem [{\citenamefont {Zener}(1932)}]{ZenerLandau}%
  \BibitemOpen
  \bibfield  {author} {\bibinfo {author} {\bibfnamefont {C.}~\bibnamefont
  {Zener}},\ }\href {\doibase 10.1098/rspa.1932.0165} {\bibfield  {journal}
  {\bibinfo  {journal} {Proceedings of the Royal Society of London A:
  Mathematical, Physical and Engineering Sciences}\ }\textbf {\bibinfo {volume}
  {137}},\ \bibinfo {pages} {696} (\bibinfo {year} {1932})}\BibitemShut
  {NoStop}%
\bibitem [{\citenamefont {Whittaker}\ and\ \citenamefont
  {Watson}(1996)}]{ModAn}%
  \BibitemOpen
  \bibfield  {author} {\bibinfo {author} {\bibfnamefont {E.~T.}\ \bibnamefont
  {Whittaker}}\ and\ \bibinfo {author} {\bibfnamefont {G.~N.}\ \bibnamefont
  {Watson}},\ }\href@noop {} {\emph {\bibinfo {title} {A Course of Modern
  Analysis}}}\ (\bibinfo  {publisher} {Cambridge University Press},\ \bibinfo
  {year} {1996})\BibitemShut {NoStop}%
\bibitem [{\citenamefont {Chen}\ \emph {et~al.}(2016)\citenamefont {Chen},
  \citenamefont {Sigrist},\ and\ \citenamefont {Schnyder}}]{WeiChen16}%
  \BibitemOpen
  \bibfield  {author} {\bibinfo {author} {\bibfnamefont {W.}~\bibnamefont
  {Chen}}, \bibinfo {author} {\bibfnamefont {M.}~\bibnamefont {Sigrist}}, \
  and\ \bibinfo {author} {\bibfnamefont {A.~P.}\ \bibnamefont {Schnyder}},\
  }\href {http://stacks.iop.org/0953-8984/28/i=36/a=365501} {\bibfield
  {journal} {\bibinfo  {journal} {Journal of Physics: Condensed Matter}\
  }\textbf {\bibinfo {volume} {28}},\ \bibinfo {pages} {365501} (\bibinfo
  {year} {2016})}\BibitemShut {NoStop}%
\bibitem [{\citenamefont {Zak}(1989)}]{Zak89}%
  \BibitemOpen
  \bibfield  {author} {\bibinfo {author} {\bibfnamefont {J.}~\bibnamefont
  {Zak}},\ }\href {\doibase 10.1103/PhysRevLett.62.2747} {\bibfield  {journal}
  {\bibinfo  {journal} {Phys. Rev. Lett.}\ }\textbf {\bibinfo {volume} {62}},\
  \bibinfo {pages} {2747} (\bibinfo {year} {1989})}\BibitemShut {NoStop}%
\bibitem [{\citenamefont {Rigol}\ \emph {et~al.}(2008)\citenamefont {Rigol},
  \citenamefont {Dunjko},\ and\ \citenamefont {Olshanii}}]{Rigol2008}%
  \BibitemOpen
  \bibfield  {author} {\bibinfo {author} {\bibfnamefont {M.}~\bibnamefont
  {Rigol}}, \bibinfo {author} {\bibfnamefont {V.}~\bibnamefont {Dunjko}}, \
  and\ \bibinfo {author} {\bibfnamefont {M.}~\bibnamefont {Olshanii}},\ }\href
  {\doibase 10.1038/nature06838} {\bibfield  {journal} {\bibinfo  {journal}
  {Nature}\ }\textbf {\bibinfo {volume} {452}},\ \bibinfo {pages} {854}
  (\bibinfo {year} {2008})}\BibitemShut {NoStop}%
\bibitem [{\citenamefont {Vidmar}\ and\ \citenamefont
  {Rigol}(2016)}]{LevRigol2016}%
  \BibitemOpen
  \bibfield  {author} {\bibinfo {author} {\bibfnamefont {L.}~\bibnamefont
  {Vidmar}}\ and\ \bibinfo {author} {\bibfnamefont {M.}~\bibnamefont {Rigol}},\
  }\href {http://stacks.iop.org/1742-5468/2016/i=6/a=064007} {\bibfield
  {journal} {\bibinfo  {journal} {Journal of Statistical Mechanics: Theory and
  Experiment}\ }\textbf {\bibinfo {volume} {2016}},\ \bibinfo {pages} {064007}
  (\bibinfo {year} {2016})}\BibitemShut {NoStop}%
\end{thebibliography}%

\end{document}